\documentclass[11pt,oneside,letterpaper]{article}
\usepackage{graphicx,amsmath,amsfonts}
\usepackage{setspace}
\usepackage{amssymb}
\usepackage{graphicx,amsmath,amsfonts}
\usepackage{caption}
\usepackage{amsmath}

\usepackage{amssymb}

\usepackage{commath}

\usepackage{amsmath}
\usepackage{amsfonts}
\usepackage{mathtools}
\usepackage{graphicx}
\usepackage{enumerate}
\usepackage{color}
\usepackage{framed}
\usepackage[margin=1in]{geometry}
\usepackage{natbib}
 \bibpunct[, ]{(}{)}{,}{a}{}{,}%
\DeclareMathOperator*{\argmax}{arg\,max}

\usepackage{natbib}
 \bibpunct[, ]{(}{)}{,}{a}{}{,}%
 \newcommand{\Br}{\linebreak[0]}
 \newcommand{\Bv}{\vec{B}}
 
\usepackage{dsfont}
\usepackage{algorithm2e}

\usepackage{graphicx,amsmath,amsfonts}
\usepackage{caption}
\usepackage{subcaption}
\usepackage{amsmath}

\usepackage{amssymb}
\usepackage{amsthm}
\usepackage{caption}
\usepackage{subcaption}

\usepackage{commath}
\usepackage{hyperref}

\newtheorem{corollary}{\textbf{Corollary}}
\newtheorem{lemma}{\textbf{Lemma}}
\newtheorem{definition}{\textbf{Definition}}
\newtheorem{theorem}{\textbf{Theorem}}
\newtheorem{example}{\textbf{Example}}
\newtheorem{assumption}{\textbf{Assumption}}

\begin{document}

\title{Maximizing Sequence-Submodular Functions and its Application to Online Advertising}
\date{}
\author{Saeed Alaei \thanks{%
Google Research, \texttt{saeed.a@gmail.com}} \and Ali Makhdoumi \thanks{%
Fuqua School of Business, Duke University, \texttt{ali.makhdoumi@duke.edu}}
\and Azarakhsh Malekian \thanks{%
Rotman School of Management, University of Toronto, \texttt{%
azarakhsh.malekian@rotman.utoronto.ca}} }
\maketitle
\begin{abstract}
Motivated by applications in online advertising, we consider a class of maximization problems where the objective is a function of the sequence of actions as well as the running duration of each action. For these problems, we introduce the concepts of \emph{sequence-submodularity} and \emph{sequence-monotonicity} which extend the notions of submodularity and monotonicity from functions defined over sets to functions defined over sequences. We establish that if the objective function is sequence-submodular and sequence-non-decreasing, then there exists a greedy
algorithm that achieves $1-1/e$ of the optimal solution. 

We apply our algorithm and analysis to two applications in online advertising: online ad allocation and query rewriting. We first show that both problems can be formulated as maximizing non-decreasing sequence-submodular functions. We then apply our framework to these two problems, leading to simple greedy approaches with guaranteed performances. In particular, for online ad allocation problem the performance of our algorithm is $1-1/e$, which matches the best known existing performance, and for query rewriting problem the performance of our algorithm is $1- 1/e^{1-1/e}$ which improves upon the best known existing performance in the literature. 
\end{abstract}




\section{Introduction}

Search advertising continues to power the growth of online advertising. For instance, Google AdWord's revenue accounts for most of its revenue, amounting to over twenty billion dollars in the first quarter of 2018.\footnote{\url{https://www.cnbc.com/2018/06/27/googles-adwords-and-doubleclick-have-been-rebranded-and-reorganized.html}} At the core of advertising for such search engines, there is a demand to allocate relevant ads to user queries. Advertisers bid on queries that are most likely to generate clicks and conversions for them, and ad allocators want relevant ads for their users to maximize revenue on their platforms. This problem can be cast as an online allocation problem where the search engine decides on the order of the ads to show for a sequence of arriving queries. The goal of the ad allocator is to maximize its revenue which is a function of the sequence of allocations.

This online allocation problem demands to develop computationally efficient solutions with guaranteed performances. In this regard, the greedy approach is a natural choice leading to simple implementations. In particular, the greedy algorithm has the advantage that it can be applied in a variety of settings where complete
knowledge of the problem is not available or in online settings where the input is revealed gradually. For maximizing set-submodular functions \cite{NWF78, NW78}, and \cite{W82} show that a greedy
algorithm achieves $1- 1/e$ of the optimal solution (subject to some constraints). However, the online ad allocation objective function is defined over sequences rather than sets for which the order of allocations matter. This raises the question of whether there exists natural extensions of set-submodular functions to functions defined over sequences. 
In this paper, we develop such a framework which enables us to obtain performance guarantees for greedy algorithms that are the same as non-decreasing submodular functions over sets. We apply our algorithm and analysis to two applications in online advertising, namely, online ad allocation and query rewriting and provide computationally efficient algorithms with guaranteed performances.

\subsection{Contribution}
The contribution of our work is twofold. First, we introduce a framework for solving a broad class of maximization problems where the objective function is defined over sequences. In particular, we introduce the notion of \emph{sequence-submodular} functions
which extends the notion of submodularity over sets to submodularity over sequences. We define sequence submodularity over both continuous and discrete sequences and carry out the analysis of these two cases separately.
Our main results show that if the objective function is sequence-submodular, \emph{sequence-non-decreasing}, and in the case of continuous sequences
\emph{differentiable}, then a greedy approach achieves $1- 1/e$ of the optimal solution subject to a constraint on the maximum length of the sequence. In our algorithm, we solve a collection of local optimization problems and then form a global solution based on the solution of these local problems. Furthermore, we show that even if the local problems cannot be solved optimally (e.g., because of limited computational or time resources), our algorithm and analysis still provide a performance guarantee. In particular, if the solutions of the local optimization problems are at least $\alpha$ times the optimal local solutions, then our algorithm achieves $1-1/e^{\alpha}$ of the overall optimal solution.

Second, we present two applications of our framework to search advertising. 
In our first application, we show that online ad allocation problem can be formulated as maximizing a sequence-submodular function. We then apply our algorithm and analysis and obtain a  greedy algorithm that achieves $\left( 1- 1/e \right) - \text{(bid to budget ratio)}$ of the optimal revenue, where the bid to budget ratio is defined as the ratio of the maximum payment to minimum budget.\footnote{We consider a general ad allocation problem in which the query distribution is unknown. Therefore, we cannot use LP rounding to solve this problem.} In particular, if the bid to budget ratio is very small, then our algorithm achieves $1- 1/e$ of the optimal revenue. This is the same as the one obtained in \cite{GM08}, using a more involved analysis based on the techniques developed in \cite{KVV90}. In our second application, we consider query rewriting for online ad allocation which is a technique to improve the relevance of ad allocation. Again, we show that ad allocation with query rewriting can be formulated as maximizing a sequence-submodular function. We then apply our algorithm and analysis and obtain a  greedy algorithm that achieves $\left( 1-1/e^{1-\frac{1}{e}} \right) - \text{(bid to budget ratio)}$ of the optimal solution. In particular, if the bid to budget ratio is very small, then our algorithm achieves $1-1/e^{1-\frac{1}{e}} \approx 0.47$ of the optimal revenue, improving upon the $1/4$ approximation of \cite{MCKW08}.



\subsection{Related Work}

Submodularity has been studied in more depth in recent years because of its applications to combinatorial auctions and the fact that many important problems in computer science, economics, and operations can be formulated as submodular function maximization. Instances include submodular welfare maximization problems such as \cite{LLN06, KLMM05, dobzinski2006improved, vondrak2008optimal, feige2009maximizing}; viral marketing and influence maximization over a network such as \cite{kempe2003maximizing}, \cite{mossel2010submodularity}; and generalized assignment problems such as \cite{FGMS06}. Another prominent application is machine learning and in particular data summarization \cite{mirzasoleiman2016distributed, lin2011class}, recommender systems \cite{gabillon2013adaptive}, and crowd teaching \cite{singla2014near}. 

The greedy approach is a natural tool to solve maximization problems with a submodular objective function. \cite{NW78} show that the greedy approach gives a $1-1/e$-approximation for maximizing a
non-decreasing submodular function over a uniform matroid. \cite{NWF78} consider this
problem over the independence system. They show that if the independence system is the intersection of $M$ matroids,
the greedy algorithm gives a $1/(M+1)$ approximation. \cite{GS07} generalize both of these
results and show that if an $\alpha$-approximate incremental oracle is available, then the greedy solution is a
$1- 1/e^{1/\alpha}$ approximation for maximizing a non-decreasing submodular function over a uniform
matroid and an $1/ (\alpha M + 1)$ approximation for the intersection of $M$ matroids. \cite{FMV07} provide a
general framework for solving the non-monotone submodular problems. More recently, \cite{asadpour2015maximizing} and \cite{adamczyk2016submodular} study maximizing stochastic submodular functions. In particular, \cite{asadpour2015maximizing} show that a greedy algorithm obtains $1/2$ of the optimal value subject to a matroid constraint. Furthermore, they prove that the greedy algorithm obtains $1 -1/e$ of the optimal value for uniform matroid constraints. Relatedly, \cite{golovin2011adaptive} extend submodularity to adaptive policies for solving stochastic optimization
problems under partial observability.

Our paper generalizes the concept of submodularity to functions defined over sequences instead of sets. 
Since the circulation of an early version of our paper, extensions of submodularity to other interesting settings have been studied in \cite{li2017inhomogeneous, tschiatschek2017selecting}, and \cite{mitrovic2018submodularity}. In particular, \cite{li2017inhomogeneous} consider a combination of submodularity and hypergraphs within the context of hypergraph clustering. More recently, \cite{tschiatschek2017selecting} and \cite{mitrovic2018submodularity} use a directed graph connecting the items together with a submodular function on its edges to define functions over sequences. In their setting, the edges of the directed graph encode the additional value of selecting elements in a particular order. Their setting and results are different from ours. In particular, the sequence functions that are defined with directed graphs do not have diminishing return property and are not equivalent to our class of sequence-submodular functions. Moreover, the guarantee of their proposed algorithm depends on the maximum degree in the underlying graph and the length of the sequence and is worse than $1- 1/e$. 

Our first application is online ad allocation problem. 
There is a considerable amount of work on AdWords auctions and in particular online ad allocation problem (see \cite{mehta2013online} for a survey). In the online ad allocation problem, the goal is to match incoming queries to advertisers with the goal of maximizing the revenue. Several papers such as \cite{MSVV07,
LPSV07} have studied this problem. In particular, assuming that the maximum bid is very small compared to budgets, \cite{MSVV07} provide a deterministic algorithm with the
competitive ratio of $1 - 1/e$ in the worst case model. It can be shown that the competitive ratio for the
greedy algorithm is $1/2$ in the worst case model. Subsequently, \cite{GM08} showed that the competitive
ratio of the greedy approach in the i.i.d model is $1 - 1/e$ and the analysis is tight. Their proof is partly based on the techniques used in \cite{KVV90} for the online
bipartite matching problem. Our framework, however, provides a simple greedy algorithm that achieves the same $1 - 1/e$ competitive ratio (under the same common assumption that the maximum bid is very small compared to budgets). The offline variant of ad allocation has been studied in \cite{AM04} and \cite{FGMS06}, where they show that the problem is NP-complete with the best known approximation factor of $1 - 1/e$.


Our second application is query rewriting in online ad allocation. There is a large literature on
clustering and mining of search logs to generate query suggestions
for improving web and paid search results. The goal of query rewriting is to define a succinct set of rewrites and assign each query to a subset of rewrites. With query rewriting, for each arriving query, the ad allocator finds the most relevant ad by searching over the ads associated with the rewrites assigned to that query (see \cite{jones2006generating, zhang2007comparing}, and \cite{singh2012rewriting}). In particular, \cite{MCKW08}, consider the problem of query rewriting in the context of search advertising and provide an algorithm that achieves $1/4$ of the optimal revenue. Again, our framework provides a simple greedy algorithm that achieves the improved $1 - 1/e^{1-\frac{1}{e}} \approx 0.47$ approximation.
\subsection{Organization}
In Section \ref{sec:AdAlloc}, we formulate the online ad allocation problem as maximizing a function defined over continuous sequences. In Section \ref{sec:model}, we introduce non-decreasing sequence-submodular functions over continuous sequences and establish that in maximizing such a function, a greedy algorithm achieves $1- 1/e$ of its optimal solution. We then show that the objective function formulating the online ad allocation problem is non-decreasing and sequence-submodular, showing that a greedy algorithm achieves $1-1/e$ of its optimal solution. In Section \ref{sec:query}, we introduce query rewriting problem in the context of search advertising and formulate it as maximizing a function defined over discrete sequences. In Section \ref{sec:Dsicrete}, we describe our framework for maximizing non-decreasing sequence-submodular functions defined over discrete sequences. We then show that the objective function formulating the query rewriting problem is non-decreasing and sequence-submodular, and establish that our greedy algorithm achieves $1-1/e^{1-1/e}$ of its optimal solution. We conclude the paper in Section \ref{sec:Conclusion}. All the omitted proofs are included in the appendix.  

\section{Online Ad Allocation}\label{sec:AdAlloc}
Search advertising constitutes one of the largest resource allocation problems, both in terms of the capital and the number of items. In online advertising mechanisms used by search engines, advertisers submit their bid to the search engine for each keyword (also referred to as \emph{query type}) and their total budget. Whenever a user searches for a service or product, the search engine (also referred to as \emph{ad allocator}) decides on a set of relevant ads to display. At the core of this service, there is an allocation algorithm that allocates ads to an arriving query, based on the relevance of the query to ads and the budget of advertisers. This problem is inherently online since the ad allocator needs to show ads whenever a query arrives, and it does not have complete information about
the arriving queries in advance. The objective of this online ad allocation problem is to find a way to perform this allocation to achieve maximum revenue. We next formally define this problem.

\subsection{Problem Formulation}

\begin{figure}[t]
\centering
    \includegraphics[width=0.35\textwidth]{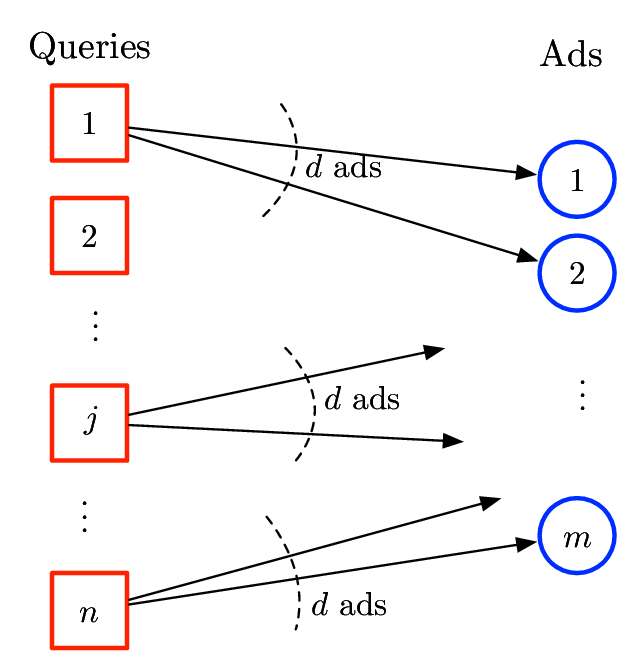}
\caption{A configuration $s$ assigns each query $j \in \mathcal{Q}=\{1, \dots, n\}$ to at most $d$ ads in $\mathcal{A}=\{1, \dots, m\}$.}
    \label{fig:Example1}
\end{figure}

We let $\mathcal{A}$ denote the set of $m$ ads (also referred to as advertisers), and $\mathcal{Q}$ denote the set of $n$ query types. When a query arrives, the ad allocator assigns this query to a set of ads to be displayed along-side organic results. To capture the limit on the number of slots for sponsored ads, we assume each query is assigned to at most $d$ ads. Each assignment of a query to an advertiser generates revenue for the ad allocator which is equal to the payment of the advertiser. In particular, we let $p_{ij}$ be the payment of advertiser $i \in \mathcal{A}$ to the ad allocator for
showing ad $i$ to a query of type $j \in \mathcal{Q}$. This payment is a function of the click-through rate of the ad, the relevance of the ad to the query, the bid of the advertiser for that query, and possibly other parameters. Each advertiser has a limited budget and showing the ad of an advertiser that has consumed her entire budget cannot provide revenue. For each advertiser $i \in \mathcal{A}$, we let $B_i \in \mathbb{R}^{+}$ denote its budget which represents the total payment the advertiser is willing to pay for a given period of time. Therefore, at each time (based on the remaining budgets of the advertisers and the payments) the ad allocator should decide on the mapping from queries to ads, defined next.
\begin{definition}[Configuration]\label{def:Config}
\textup{
A \emph{configuration} $s$ is a mapping from query types to ads such that each query type is mapped
to at most $d$ ads. Formally, a configuration $s$ is a collection of sets $s(j) \subseteq \mathcal{A}$ for all $j \in \mathcal{Q}$ such that $|s(j)| \le d$. We let $\mathcal{S}$ be the set of all possible configurations. Figure \ref{fig:Example1} illustrates the definition of configuration.
} 
\end{definition}
The revenue of the ad allocator is the sum of the revenues generated for each query given the configuration that is used by the ad allocator, as described next. 

\begin{figure}[t]
\centering
    \includegraphics[width=0.6\textwidth]{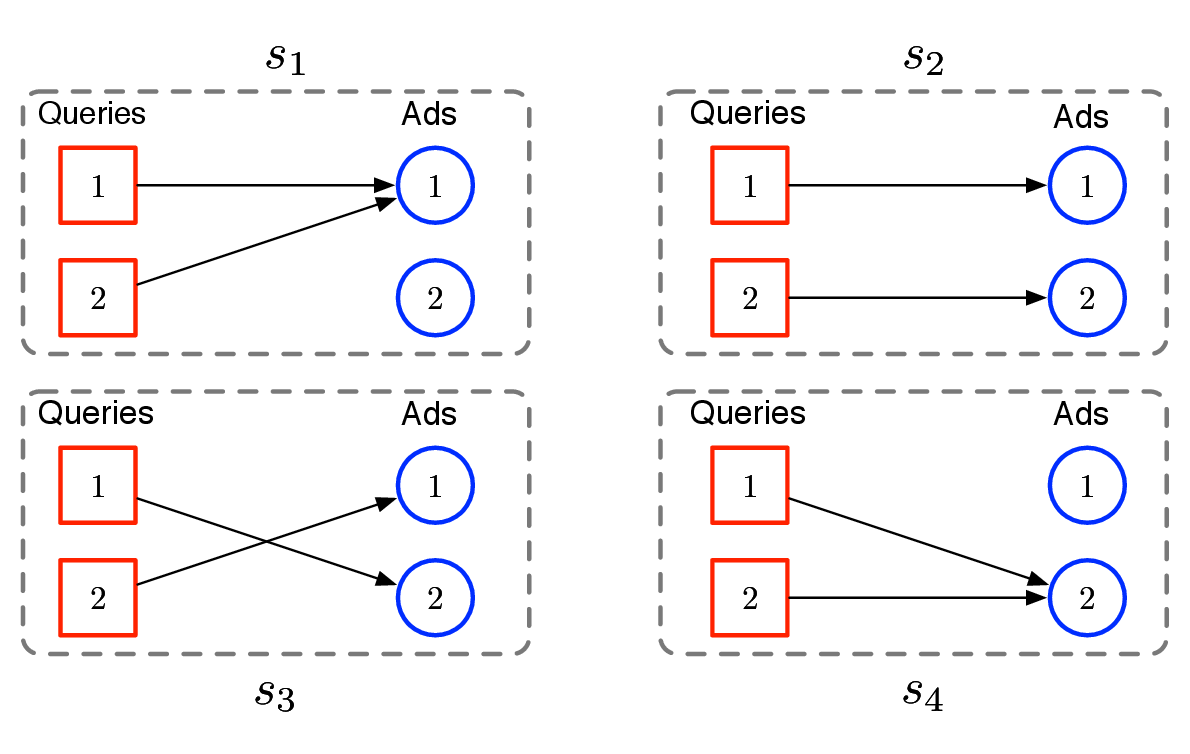}
\caption{The four configurations in the setting of Example \ref{example:AdAlloc2}.}
    \label{fig:Example_1}
\end{figure}

\subsection{Ad allocator's Decision and Revenue}

We let $T$ denote the end of time horizon and assume a sequence of queries are arriving over $[0,T]$ according to a Poisson point process with rate $1$ (the rate one assumption has no bearing on our results, but simplifies the exposition). The type of each query is an i.i.d. random variable drawn from a fixed but
possibly unknown distribution $\mathbf{q}=(q_1, \cdots, q_n)$ where $q_j$ is the probability of a query being of type $j$. Therefore, for any interval of length $\Delta t \in \mathbb{R}^{+}$, 
 the expected number of type $j$ queries arriving in a period of length $\Delta t$ is $(\Delta t) q_j$.

The next definition captures the decision of ad allocator regarding the sequence of configurations used during the period $[0,T]$. 
\begin{definition}[Allocation Strategy]\label{def:AllocStra}
We call any sequence of configurations over time $[0, T]$ an \emph{allocation strategy} which is represented by 
\begin{align*}
	H=\left( \left(s_1, \Delta t_1 \right), \cdots, \left( s_k, \Delta t_k \right) \right),
\end{align*}
where $s_i \in \mathcal{S}$, $\Delta t_i \in \mathbb{R}^+$, $\sum_{i=1}^{k} \Delta t_k=T$, and $k \in \mathbb{N}$. This sequence indicates that the ad allocator uses
each configuration $s_i$ (in order) for a duration of $\Delta t_i$ for all $i \in \{1, \cdots, k\}$. We let $\mathds{H}(\mathcal{S})$ denote the set of all possible allocation strategies. 
\end{definition}


 For any given allocation strategy $H$ we let $u(H)$ be the expected revenue of the ad allocator for using allocation strategy $H$, where the expectation is taken with respect to the Poisson random process governing the arrival of queries and distribution $\mathbf{q}$ for the type of queries. Therefore, the problem of the ad allocator can be written as 
\begin{align*}
\max_{H \in \mathds{H}(\mathcal{S})}  & ~u(H).
\end{align*}
Note that the sequence $H$ can be chosen adaptively. That is, for any $t \in [0, T]$, the configuration used at time $t$ depends on the query types that have arrived before time $t$ as well as the configurations used to serve those queries.\footnote{More precisely, if for any $s \in [0,T]$ we let $\mathcal{F}_s$ be the $\sigma$-algebra of the events happened in the interval $[0, s)$, then the configuration used at time $t$ must be $\mathcal{F}_t$-measurable.}
In the next example, we illustrate the definition of allocation strategy and revenue of the ad allocator. 

\begin{example}\label{example:AdAlloc2}
Suppose $d=1$ and there are two advertisers and two query types, i.e., $\mathcal{A}=\{1,2\}$ and $\mathcal{Q}=\{1, 2\}$. We let $T=1$ and the payments be 
\begin{align*}
p_{11}=2, p_{12}=1, p_{21}=1, p_{22}=3,
\end{align*}
and the budgets be $B_1$ and $B_2$. We also let $q_1$ and $q_2$ be the probability of query types $1$ and $2$, respectively. In this setting, there exist four possible configurations as follows (see Figure \ref{fig:Example_1})
\begin{align*}
s_{1}& = \left( s_1(1)=\{1\}, s_1(2)=\{1\} \right), \qquad \quad  s_{2}= \left( s_2(1)=\{1\}, s_2(2)=\{2\} \right),\\
s_{3}& = \left( s_3(1)=\{2\}, s_3(2)=\{1\} \right), \qquad \quad s_{4}= \left( s_4(1)=\{2\}, s_4(2)=\{2\} \right).
\end{align*}
For instance, configuration $s_1$ maps both queries to ad $1$ and configuration $s_2$ maps queries of type $1$ to ad $1$ and queries of type $2$ to ad $2$. 
 We next find the revenue of the allocation strategy $A=((s_1, 1))$, i.e., running configuration $1$ for duration $[0,1]$. Since $s_1(1)=s_1(2)=\{1\}$, for both types of queries the ad allocator shows ad $1$ until ad $1$ runs out of budget. We let $N_1$ be the number of type one queries in period $[0,1]$ and $N_2$ be the number of type two queries. For a given $N_1$ and $N_2$, to find the revenue we need to consider the following cases:
\begin{itemize} 
\item  The budget constraint is not binding, i.e., $N_1 p_{11}+ N_2 p_{12} \le B_1$: in this case, we can show ad $1$ for both types of queries for the entire time interval. The generated revenue in this case becomes $N_1 p_{11}+ N_2 p_{12}$. 
\item The budget constraint is binding, i.e., $N_1 p_{11}+ N_2 p_{12} > B_1$: in this case we can only show ad $1$ so long as its remaining budget is above the payment for the query type. In this case, the revenue belong to the interval $(B_1- \max_{j \in \mathcal{Q}}p_{1j},B_1]$. 
\end{itemize}  
\end{example}
\section{Continuous Sequence-Submodular Functions and their Maximization}
    \label{sec:model}

In this section, we define our framework for sequence submodular function maximization and then apply it to online ad-allocation problem. To this end, we first introduce continuous sequences together with the notion of sequence functions, their monotonicity, and their submodularity. We then find a greedy algorithm to maximize sequence submodular functions and establish its performance. Finally, we apply our algorithm and analysis to online ad allocation problem. 

\subsection{Continuous Sequences: Definition and Operations}

For any finite set of elements $\mathcal{S}$, the sequence $A=\left( (s_1, \Delta t_1),
\cdots, (s_k, \Delta t_k) \right)$ where $k \in \mathbb{N}\cup\{0\}$ and $s_i \in \mathcal{S}$ and $\Delta t_i \in \mathbb{R}^+$ is
called a \emph{continuous sequence}. The length of a continuous sequence $A=((s_1, \Delta t_1), \cdots, (s_k,
\Delta t_k))$ denoted by $|A|$ is equal to $\sum_{i=1}^k \Delta t_i$. We denote the set of all finite continuous sequences of
$\mathcal{S}$ by $\mathds{H}^C(\mathcal{S})$, formally defined as

\begin{align*}
    \mathds{H}^C(\mathcal{S})  = \left\{A = \left( (s_1, \Delta t_1), \cdots, (s_k, \Delta t_k) \right) ~|~  k \in \mathbb{N}\cup\{0\}, s_i \in \mathcal{S}, \Delta t_i \in \mathbb{R}^+ \right\}.
\end{align*}
Throughout, we use $\emptyset$ to denote the empty sequence. For instance, each allocation strategy in online ad allocation problem is a continuous sequence where each configuration (see Definition \ref{def:Config}) is an element and each allocation strategy (see Definition \ref{def:AllocStra}) is a continuous sequence. 



We say two continuous sequences $A$ and $B$ are \emph{equivalent} and denote it by $A \equiv B$
if they have the same length and their corresponding elements are the same. We next define three key operations on continuous sequences which we use throughout our analysis. 
\begin{definition}[Concatenation]
For two continuous sequences $A$ and $B$, their \emph{concatenation} denoted by $A \bot B$ is a new sequence that is the result of attaching the beginning of sequence $B$ to the end of sequence $A$. 
\end{definition}
For instance, in the context of online ad allocation the concatenation of two allocation strategies $A$ and $B$, i.e., $A \bot B$ is an allocation strategy that uses the configurations specified by $A$ (in order) followed by configurations specified by $B$. 
\begin{definition}[Refinement]
The \emph{refinement} of a continuous sequence $A=\left((s_1, \Delta t_1), \cdots, (s_k, \Delta t_k) \right)$ in the interval $[x,y)$ denoted by $A_{[x, y)}$ is a subsequence of $A$ that contains all the elements of $A$ starting from time $x$ to time $y$. Formally, we have 
\begin{align*}
    A_{[x,y)}  = \left( (s_f, \Delta t_f-\delta), (s_{f+1}, \Delta t_{f+1}), \cdots, (s_{l-1}, \Delta t_{l-1}), (s_l, \Delta t_l-\delta') \right),
\end{align*}
where $f,l \in \mathbb{N}$ and $\delta,\delta' \in \mathbb{R}^+\cup\{0\}$ are uniquely defined from the following relations:
\begin{align*}
    \sum_{i=1}^{f-1} \Delta t_i \le x < \sum_{i=1}^f \Delta t_i, ~ \delta = x-\sum_{i=1}^{f-1} \Delta t_i, \quad \sum_{i=1}^{l-1} \Delta t_i  < y \le \sum_{i=1}^l \Delta t_i,  ~  \delta'  = \sum_{i=1}^l \Delta t_i-y.
\end{align*}
\end{definition}
For instance, in the context of online ad allocation, the refinement of allocation strategy $A$ in the interval $[x, y)$ is an allocation strategy which contains all the configurations specified by $A$ (in the same order) from time $x$ to time $y$. 
\begin{definition}[Domination]
A continuous sequence $A$ is \emph{dominated} by another continuous sequence $B$,  denoted by $A \prec B$, if we can remove some elements of $B$ to obtain $A$. Formally, for two continuous sequences $A$ and $B$, we have $A \prec B$ if and only if there exists $m \in \mathbb{N}$ and $0 \leq x_1 < x_2
< \cdots < x_{2m} \leq |B|$ such that
\begin{align*}
    A \equiv B_{[x_1, x_2)} \bot \cdots \bot B_{[x_{2m-1}, x_{2m})}.
\end{align*}
\end{definition}
For instance, in the context of online ad allocation for two allocation strategies $A$ and $B$ we have $A \prec B$, if we can obtain $A$ be removing some of the configurations in $B$ (by either decreasing their running time or eliminating the configuration entirely) and keeping the order of the remaining configurations. 

In the next section, we use these three operations to define the class of submodular non-decreasing continuous sequence functions. 

\subsection{Submodular Non-decreasing Continuous Sequence Functions}
    \label{sec:conds}

\emph{Continuous sequence functions} are functions whose domain are continuous sequences. Formally, given a finite set $\mathcal{S}$, any function $u : \mathds{H}^C(\mathcal{S}) \rightarrow \mathbb{R}$ is a continuous sequence function. We next define the key attributes of continuous sequence functions, namely sequence-non-decreasing, sequence-submodularity, and differentiability. These attributes are the analogy of non-decreasing submodular functions defined over sets and enable us to provide performance guarantees for greedy algorithms in maximizing continuous sequence functions.

\begin{definition}
    \label{cond:mono,ncomp}%
A continuous sequence function $u$ is \emph{sequence-non-decreasing} if 
\begin{align}
    \label{eq:mono}%
    & u(A) \le u(B), \quad \forall A, B \in \mathds{H}^C(\mathcal{S}) \text{ such that } A \prec B,  \nonumber \\
    & u(\emptyset) = 0.
\end{align}
A continuous sequence function $u$ is \emph{sequence-submodular} if
\begin{equation}\label{eq:ncomp}
u(C|A) \ge u(C|B), \quad  \forall A, B, C \in \mathds{H}^C(\mathcal{S}) \text{ such that } A \prec B, 
\end{equation}
where $u(B|A)$ is the \emph{marginal value} of the sequence function defined as 
\begin{align*}
	u(B|A) = u(A \bot B)-u(A), \quad  \forall A,B \in \mathds{H}^C(\mathcal{S}). 
\end{align*}
A continuous sequence
function $u : \mathds{H}^C(\mathcal{S}) \rightarrow \mathbb{R}$ is \emph{differentiable} if for any $A \in
\mathds{H}^C(\mathcal{S})$, $u(A_{[0, t)})$ is continuous and differentiable with a continuous derivative with respect to $t$ for
all $t \in [0, \infty)$ except at a finite number of points for which it may have different left and right derivatives and hence
a non-continuous derivative.
\end{definition}

In the next section, we provide a greedy algorithm to maximize any continuous sequence function satisfying the following assumption.

\begin{assumption}\label{Assump:SS+ND+Diff}
The continuous sequence function $u$ is sequence-non-decreasing, sequence-submodular, and differentiable. 
\end{assumption}

\subsection{Greedy Algorithm for Maximizing Continuous Sequence Functions} \label{sec:MaximizationCont}
In this section, we consider the problem of maximizing a continuous sequence function subject to a given length constraint. In particular, we develop a greedy algorithm for such maximization problem and establish its performance guarantee for sequence functions satisfying Assumption \ref{Assump:SS+ND+Diff}.

For a given $\mathcal{S}$ and continuous sequence function $u : \mathds{H}^C(\mathcal{S}) \rightarrow \mathbb{R}$, and time horizon $T \in
\mathbb{R}^+$, the objective is to find a sequence $H \in \mathds{H}^C(\mathcal{S})$ that maximizes $u$ subject to the length constraint $|H| \le T$, i.e., 
\begin{align}\label{eq:ContinuousCase}
\max_{H \in \mathds{H}^C(\mathcal{S})} & u(H)  \nonumber \\
\text{ s.t. }& |H| \le T.
\end{align}

We next introduce a notation that we use in stating our algorithm and results for the continuous setting. For an element $s \in \mathcal{S}$, time duration $\delta \in \mathbb{R}^+$, and continuous sequence $A \in \mathds{H}^C(\mathcal{S})$, we define $\dot{u}_s(\delta|A)$ as 

\begin{align}
    \label{eq:con:udot}
    \dot{u}_s(\delta|A) & = \frac{d}{d x}u((s, x)|A) \Bigr |_{x=\delta} 
\end{align}
which represents the rate of increasing $u(A)$ if we continue using element $s$ after using it for duration $\delta$. We also define 
\begin{equation}
    \label{eq:con:udot:last}
    \dot{u}_s(0|A) = \lim_{\delta \rightarrow 0^+} \dot{u}_s(\delta|A),
\end{equation}
which represents the rate of increasing $u(A)$ if we start using element $s$ after using sequence $A$. Note that $\dot{u}_s(\delta|A)$ is always defined (except for finite number of points) because \eqref{eq:con:udot} can be rewritten as follows
\begin{align*}
\frac{d}{d x} u((s, x)|A)  \Bigr |_{x=\delta} & = \frac{d}{d x}\left(u(A \bot (s, x)) - u(A)\right)   \Bigr |_{x=\delta} = \frac{d}{d x}u(A \bot (s, x))  \Bigr |_{x=\delta}
      = \frac{d}{d x} u((A \bot (s, \infty))_{[0,|A|+x)}) \Bigr |_{x=\delta},
\end{align*}
and by Assumption \ref{Assump:SS+ND+Diff} (in particular, differentiability) $\frac{d}{d x} u((A \bot (s, \infty))_{[0,|A|+x)}) \Bigr |_{x=\delta}$ exists except for a finite number of points. 
 Also note that
with Assumption \ref{Assump:SS+ND+Diff}, $\dot{u}_s(\delta | A)$ is a continuous function over $\mathbb{R}^+$ except at a finite number of
points.

%

Our key result presented next establishes the performance of a greedy algorithm for maximizing continuous sequence functions. 

\begin{theorem}
    \label{thm:con}
Suppose Assumption \ref{Assump:SS+ND+Diff} holds for a continuous sequence function $u$. For any $\alpha \in [0, 1]$ and a sequence $H=\left( \left(s_1, \Delta t_1\right), \cdots, \left(s_k, \Delta t_k\right) \right)$ in $\mathds{H}^{C}(\mathcal{S})$ with $|H|=T$, if for all  $t \in [0, T)$ we have 
\begin{equation}
    \label{eq:thm:con}
     \frac{d}{dt}u\left(H_{[0,t)}\right) \ge \alpha \max_{s \in \mathcal{S}} \dot{u}_s\left( 0|H_{[0,t)} \right),
\end{equation}
then
\begin{align*}
    \frac{u\left( H \right)}{u\left( O \right)} \ge 1-\frac{1}{e^\alpha},
\end{align*}
where $O \in \mathds{H}^{C}(\mathcal{S})$ denotes the optimal solution of problem \eqref{eq:ContinuousCase}.\footnote{Note that an optimal solution $O$ exists. This follows from Weierstrass extreme value theorem.}
\end{theorem}
Theorem \ref{thm:con} states that if the elements of the sequence $H$ are chosen such that at each point $t \in [0, T)$,
the derivative of $u$ is at least $\alpha$ times its optimal local maximum, then the resulting sequence yields $1- 1/e^{\alpha}$ of the optimal solution (global maximum).

We next outline the key idea of this result for $\alpha=1$ (the complete proof is given in the Appendix). First, using sequence-submodularity and differentiability, we show that the rate of increase in the function value with the greedy choice is as large as the time average marginal increase by concatenating any other sequence. Formally, for all $B \in \mathds{H}^C(\mathcal{S})$ and $t \in [0, T]$ we show  
\begin{align*}
\max_{s \in \mathcal{S}} \dot{u}_s\left( 0|H_{[0,t)} \right) \ge \frac{1}{|B|} u(B | H_{[0,t)} ).
\end{align*}
Substituting the optimal sequence, i.e., $O$, for $B$ and then using non-decreasing property, we show that this rate of increase is as large as the time average difference between the function value of the optimal solution and the function value of the current sequence. Formally, we have  
\begin{align*}
\max_{s \in \mathcal{S}} \dot{u}_s\left( 0|H_{[0,t)} \right) \ge \frac{1}{T} \left( u(O ) - u(H_{[0,t)} ) \right).
\end{align*}
This provides a recursive relation between the utility of the greedy choice and the optimal choice. Using this recursive relation we then establish that the function value of sequence $H$ is at least $1- 1/e$ times the function value of the optimal solution $O$. 


Motivated by Theorem \ref{thm:con}, Algorithm \ref{alg:cont} presents our greedy algorithm for maximizing non-decreasing, submodular, and differentiable sequence functions. 
\begin{algorithm}[t]
    \caption{Greedy algorithm for continuous setting \label{alg:cont}}
    $t \leftarrow 0$ \;
    $i \leftarrow 1$ \;
    $H \leftarrow \emptyset$ \;
    \While{$t < T$}{
        find $(s_i, \Delta t_i)$ such that $ \forall \delta \in [0, \Delta t):
            \ \dot{u}_{s_i}\left( 0|H\bot(s_i, \delta) \right) \ge \alpha \max_{s \in \mathcal{S}} \dot{u}_s\left( 0|H\bot(s,\delta) \right)$ \;
        $H \leftarrow H \bot (s_i, \Delta t_i)$ \;
        $t \leftarrow t+\Delta t_i$ \;
        $i \leftarrow i+1$ \;
    }
\end{algorithm}

Algorithm \ref{alg:cont} starts with an empty sequence $H$ (the initialization is $H=\emptyset$) and at each time $t$ finds an element $s_i$ together with an interval of running it (i.e., $[t, t+\Delta t_i)$) such that at any time in this interval, the rate of increasing function $u$ is at least $\alpha$ times the maximum rate of increase among all elements of $\mathcal{S}$. The following which is an immediate corollary of Theorem \ref{thm:con} formally states the performance of Algorithm
\ref{alg:cont}.


\begin{corollary}
    \label{thm:con:fin}
Suppose Assumption \ref{Assump:SS+ND+Diff} holds for a continuous sequence function $u$. Algorithm
\ref{alg:cont} generates a continuous sequence with value at least $1-1/e^\alpha$ of the optimal solution. 
\end{corollary}

We point out a few remarks regarding Algorithm \ref{alg:cont}. First, note that the algorithm, in general, may not terminate, however, if it terminates with the resulting $H$, then $u(H)$ is at least $1-\frac{1}{e^\alpha}$ times the optimal solution. Second, in Algorithm \ref{alg:cont}, each time that we switch the element in use, we need an \emph{incremental oracle} to find the next element and the duration of using it which is $\alpha$-optimal. This incremental oracle is specific to each problem. We next show that for online ad allocation problem, the algorithm terminates in finite time and the incremental oracle can be found exactly, i.e., with $\alpha=1$.

\subsection{Application to Online Ad Allocation}\label{sec:ContOptForAdAlloc}
We first introduce a slight variation of the online ad allocation problem and then show the revenue function for that variation satisfies Assumption \ref{Assump:SS+ND+Diff}. We then use Theorem \ref{thm:con} to establish the performance of our greedy algorithm for the variation. Finally, using these results, we establish the performance of the greedy algorithm for the original online ad allocation problem.

Recall that $p_{ij}$ is the payment of advertiser $i \in \mathcal{A}$ to the ad allocator for
showing ad $i$ to a query of type $j \in \mathcal{Q}$. Also, the type of each query is an i.i.d. random variable drawn from a fixed but
possibly unknown distribution $\mathbf{q}=(q_1, \cdots, q_n)$ where $q_j$ is the probability of a query being of type $j$. Also, recall that $\mathcal{S}$ is the set of configurations for online ad allocation problem, $\mathds{H}(\mathcal{S})$ is the set of all allocation strategies, and $u: \mathds{H}(\mathcal{S}) \to \mathbb{R}$ is a function that maps an allocation strategy to its expected utility.

We consider a variation of online ad allocation problem in which whenever an advertiser runs out of budget (i.e., its budget is less than the payment of the ad specified by the configuration in use) the ad allocator shows the ad for a fraction of time and charges the advertiser for that fraction. 
We let $\tilde{u}: \mathds{H}(\mathcal{S}) \to \mathbb{R}$ be the function that maps an allocation strategy to its expected utility in this variation of online ad allocation problem.


In the next lemma we show that the utility function of the variation of online ad allocation problem satisfies Assumption \ref{Assump:SS+ND+Diff}. 
\begin{lemma}\label{thm:onlineAd}
The expected revenue of the variation of online ad allocation problem , i.e., $\tilde{u}: \mathds{H}(\mathcal{S}) \to \mathbb{R}$ is sequence-non-decreasing, sequence-submodular, and differentiable. 
\end{lemma}
We next outline the idea to prove sequence-submodularity of the variation of online ad allocation problem (the complete proof of all properties is given in the appendix). We next consider the allocation strategies $A,B,C \in \mathds{H}(\mathcal{S})$ with $A \prec B$ and show $\tilde{u}(C | A) \ge \tilde{u}(C|B) $. The idea is to compare the contribution of each ad to $\tilde{u}(C | A)$ and $\tilde{u}(C|B)$. We first show that the remaining budget of each advertiser after $A$ is greater than (or equal to) its budget after running $B$. We then divide ads into two categories: 
\begin{itemize}
\item Ads that have exhausted all of their budget after running $A\bot C$. The contribution of these ads to $\tilde{u}(C | A)$ is all their remaining budgets after running $A$. On the other hand, the contribution of these ads to $\tilde{u}(C | B)$ is at most their remaining budget after running $B$, which is smaller than their contribution to $\tilde{u}(C | A)$. 

\item Ads that still have budget after running $A \bot C$. Since these ads do not run out of budget, the allocation strategy $C$ (when running after $A$) has extracted revenue from these ads at the full rate. 
\end{itemize}

Lemma \ref{thm:onlineAd} together with Theorem \ref{thm:con} establish the performance guarantee of an allocation strategy obtained by using greedy algorithm for the variation of the online ad allocation problem. Using this result, we next show the performance guarantee of an allocation strategy obtained by using greedy algorithm for the original online ad allocation problem. 


\begin{lemma}\label{Lem:jaDd_1}
Let $H$ be the allocation strategy obtained by running greedy algorithm for the original online ad allocation problem. We have 
\begin{align*}
u(H) \ge \left( \left(1- \frac{1}{e} \right) -  \left( \frac{\max_{i \in \mathcal{A}, j \in \mathcal{Q} } p_{ij}}{ \min_{i \in \mathcal{A}} B_i} \right) \right) u(O),
\end{align*}
where $O$ is the optimal allocation strategy. 
\end{lemma}
This lemma establishes the performance guarantee of an allocation strategy obtained by running Algorithm \ref{alg:cont}, i.e., by greedily choosing the configuration with the highest rate of revenue increase (equivalently, finding $s$ such that Eq. \eqref{eq:thm:con} holds with $\alpha=1$). To find such a configuration, we need to compute $\dot{u}_s(0|H_{[0,t)})$, which we find in the next lemma.

\begin{lemma}\label{Lem:DerivativeOnlineAdAlloc}
Let $\mathcal{S}$ be the set of configurations for online ad-allocation problem and let $H=((s_1, \Delta t_1), \dots, (s_k, \Delta t_k))$ be an allocation strategy. For any $t$ and $s \in \mathcal{S}$, we have 
\begin{align*}
    \dot{u}_s(0|H_{[0,t)})= \sum_{j \in \mathcal{Q}} q_j	\sum_{i \in s(j)} p_{ij},
\end{align*}
assuming that all ads specified by $s$ have budget. 
\end{lemma}
This lemma holds because for a Poisson point process with rate $1$ as $\delta \to 0$ the probability of having more than one query in an interval of length $\delta$ is $O(\delta^2)$ and the probability of having one query is $\delta$. Therefore, as $\delta \to 0$, $\dot{u}_s(\delta|H_{[0,t)})$ becomes the expected revenue generated by one arriving query. Since the arriving query is of type $j$ with probability $q_j$,  the expected increase in the revenue becomes $\sum_{j \in \mathcal{Q}} q_j \sum_{i \in s(j)} p_{ij}$. 

Using Lemma \ref{Lem:DerivativeOnlineAdAlloc}, if at time $t$ the ad allocator uses configuration $s$ such that for all $j \in \mathcal{Q}$
\begin{align}\label{eq:Algtemp1}
s(j) \in  \argmax_{A \subseteq \mathcal{A}, ~ |A| \le d} \sum_{i \in A} p_{ij},
\end{align}
where $\mathcal{A}$ is the set of ads which still have budget, then we have 
\begin{align*}
     \frac{d}{dt}u\left(H_{[0,t)}\right) \ge \max_{s \in \mathcal{S}} \dot{u}_s\left( 0|H_{[0,t)} \right).
\end{align*}
Also note that we can keep using a configuration until at least one of the ads runs out of budget. We then update the set $\mathcal{A}$ to become the set of ads which still have budget and recompute the best configuration as given in Eq. \eqref{eq:Algtemp1}. The complete specification of Algorithm \ref{alg:cont} to online ad allocation problem is given in Algorithm \ref{alg:adallocation}.

\begin{algorithm}[t]
    \caption{Ad Allocation Algorithm \label{alg:adallocation}}
    $\Bv \leftarrow (B_1, \cdots, B_m)$ \;
    \While{$t < T$ \text{and} $\mathcal{A} \neq \emptyset$}{
        \tcp{find the best configuration}
        \For{$j \in \mathcal{Q}$}{
            $s(j) \leftarrow \argmax_{A \subseteq \mathcal{A}, ~ |A| \le d} \sum_{i \in \mathcal{A}} p_{ij}$ \;
            }
            Use configuration $s$ and keep updating $\Bv$ and $t$, until either $t \ge T$ or at least one ad runs out of budget\;
				$\mathcal{A} \leftarrow \mathcal{A} \setminus \{i~:~i \text{ is out of budget}\}$
    }
\end{algorithm}
Note that this algorithm does not require the knowledge of the distribution $\mathbf{q}$. This is because as shown in Eq. \eqref{eq:Algtemp1}, we can find the best configuration without knowing $\mathbf{q}$. 
The following proposition formally states the performance of Algorithm \ref{alg:adallocation} for solving online ad allocation problem.

\begin{theorem}\label{pro:OnlineAdAlloc}
Greedy Algorithm \ref{alg:adallocation} finds an allocation strategy whose expected revenue is at least
\begin{align*}
\left( 1-\frac{1}{e} \right) - \left( \frac{\max_{i \in \mathcal{A}, j \in \mathcal{Q} } p_{ij}}{ \min_{i \in \mathcal{A}} B_i} \right)
\end{align*}
of the optimal expected revenue.
\end{theorem}
Theorem \ref{pro:OnlineAdAlloc} directly follows from Lemmas \ref{thm:onlineAd}, \ref{Lem:jaDd_1}, and \ref{Lem:DerivativeOnlineAdAlloc} and establishes the performance guarantee of our algorithm which holds for any bid to budget ratio. As a corollary of this proposition, if the bid to budget ratio is very small, then our algorithm achieves $1- 1/e$ of the optimal solution. This approximation factor is also shown by \cite{GM08} using an involved analysis based on the techniques of \cite{KVV90}, under the same assumption that the bid to budget ratio is very small. 

\section{Query Rewriting}\label{sec:query}

In search advertising, advertisers bid on queries that are most likely to generate clicks and conversions for them. Ad allocators then match advertisers to user queries that are relevant to the advertisers. However, one issue is that a relevant ad for a given query may not necessarily exist among the set of ads that have bid for that query. Indeed, that set may be empty even though a relevant ad exists. For instance, an ad bidding on the keyword ``wedding band'' may be appropriate for the query ``engagement ring''. Another issue is that the advertiser bidding strategy for queries is ever-changing. Indeed, advertisers manage their budget throughout a given period by turning on and off their ads (either automatically or manually). This demands a system that can swiftly adapt to these changes by recomputing the keyword-ad relevance (e.g., with a machine-learned relevance ranking model), incorporating the changes in the availability of advertisers. Therefore, it is more practical to associate queries to a few keywords and then add (or remove) the list of advertisers of those keywords (the ones they are bidding on) rather than hundreds or even thousands of queries.

To address these issues, a common mechanism for search engines is \emph{query rewriting}. At the high level, query rewriting outputs a list of keywords, referred to as rewrites, that are relevant for queries in the
original list. 

\subsection{Problem Formulation}
In query rewriting, the search engine associates each query with a set of rewrites. When a query arrives, the ad allocator assigns this query to a set of ads that have bid for  at least one of the rewrites associated with that query. Therefore, the query rewriting problem becomes an online ad allocation problem with the constraint that only ads that have bid on the relevant rewrites to an incoming query can be displayed. 

Formally, we denote the set of rewrites with $\mathcal{R}$. Each rewrite $r \in \mathcal{R}$ is relevant for a subset of advertisers, i.e., only the subset of advertisers bidding on rewrite $r$, denoted by $W_r \subseteq \mathcal{A}$.  The search engine associates each query type $j \in \mathcal{Q}$ with a subset of rewrites denoted by $Y_j \subseteq \mathcal{R}$. For an incoming query of type $j \in \mathcal{Q}$, the ad allocator decides on $d$ ads from the set $\bigcup_{r \in Y_j} W_r$ to display.  For instance, the query ``engagement ring'' can be associated with rewrite ``wedding band''. Now if the advertisers in  ``wedding industry'' have bid on ``wedding band'', their ad can be shown for the query ``engagement ring''. Note that too many rewrites for a given query will slow down the time needed to serve an ad. To address this issue, we restrict the set of associated rewrites with each query type to have at most $k$ rewrites, i.e., $|Y_j| \le k$ for all $j \in \mathcal{Q}$. Therefore, the query rewriting problem is how to find a set $Y_j$ for all $j \in \mathcal{Q}$ such that $|Y_j| \le k$ together with the corresponding allocation strategy to maximize the revenue of the ad allocator. 

We suppose that in the time interval $[0,T]$ there are $T q_j$ incoming queries of type $j$, where $\mathbf{q}=(q_1, \cdots, q_n)$ is known. The order of the incoming queries, however, is unknown and random. This is a common assumption in practice because search engines have access to historical data and can estimate the distribution of the queries. It is also the same assumption as the one used in \cite{MCKW08} and \citet{mehta2013online}. 

\begin{figure}[t]
\centering
    \includegraphics[width=0.5\textwidth]{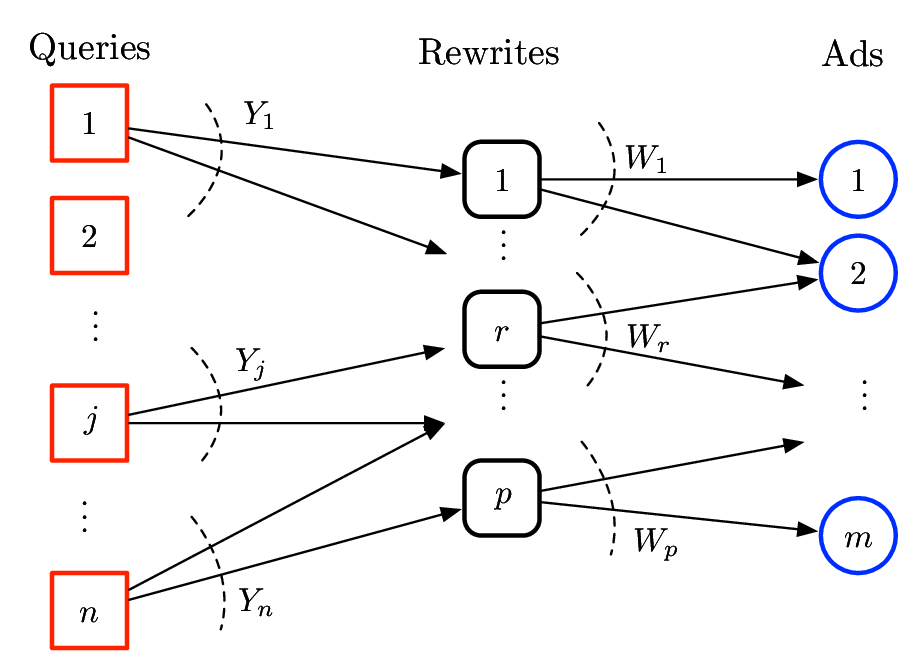}
\caption{Given the sets $Y_1, \dots, Y_{n}$, a query rewriting configuration $s^q$ assigns each query $j \in \mathcal{Q}$ to at most $d$ ads in the set $\bigcup_{r\in Y_j} W_{r}$. }
    \label{fig:query}
\end{figure}

Similar to the online ad allocation problem, given the set of rewrites, at each time the ad allocator should decide on the mapping from queries to ads, defined next.
\begin{definition}[Query Rewriting Configuration]
For a given $Y_1, \dots, Y_n$, denoting the set of rewrites associated with each query, a \emph{query rewriting configuration}, denoted by $s^{q}$, maps a given query $j$ to at most $d$ ads among those included in the set of rewrites associated with query $j$. Formally, $s^{q}$ is a collection of sets $s^q(j) \subseteq \bigcup_{r \in Y_j} W_r$ such that $|s^q(j)| \le d$ for all $j \in \mathcal{Q}$ (see Figure \ref{fig:query} for an illustration of a query rewriting configuration). We let $\mathcal{S}^q$ be the set of all such configurations. For a given sequence of queries and configurations the revenue of the ad allocator is the sum of the revenues generated for each query. We can represent the allocation of ads to queries over time $[0,T]$ by a sequence
\begin{align*}
	H^q=\left( s^q_1,\cdots, s^q_T \right), \quad  s^q_i \in \mathcal{S}^q,  i =1, \dots, T.
\end{align*}
This sequence indicates that the ad allocator uses
 configuration $s^q_i$ for the $i$th query. We call
$H^q$ a \emph{query rewriting allocation strategy} and let $\mathds{H}(\mathcal{S}^q)$ denote the set of all such strategies. 
\end{definition}


We also let $u(H^q)$ denote the expected revenue of the ad allocator for using strategy $H^q$ where the expectation is taken over all possible permutations of $T q_j$ queries of type $j$ for all $j \in \mathcal{Q}$. The query rewriting problem is to choose the set of rewrites $Y_1, \dots, Y_n$ (before the queries arrive) together with a query rewriting allocation strategy $H^q$ to maximize $u(H^q)$, i.e., 
 
\begin{align}\label{eq:QR_1}
\max_{Y_{1}, \dots, Y_{n}, H^q \in \mathds{H}(\mathcal{S}^q)}  & ~u(H^q)\\
\text{ s.t. } & ~ |Y_{j}| \le k,~ \forall j \in \mathcal{Q}. \nonumber
\end{align}
Note that the sequence $H^q$ can be chosen in an adaptive way. In particular, the $i$th configuration (i.e., $s^q_i$) can depend on the query types that have arrived before time $i$ as well as the configurations used to serve those queries (i.e., $s^q_1, \dots, s^q_{i-1}$). However, the set of rewrites $Y_1, \dots, Y_n$ are chosen in an offline fashion before the allocation of ads to queries start.

To specify a query rewriting allocation strategy $H^q$, we need to specify the configuration used at any time. We next define a discrete sequence which provides a compact representation of the query rewriting allocation strategy.  
Any query rewriting allocation strategy to serve a sequence of arriving queries determines the consumed budget that each advertiser allocates to each query type. We let $B_{ij}$ be the expected consumed budget of advertiser $i$ in serving query type $j$. We also let $\Bv^j = \left( B_{1j},
\cdots, B_{mj} \right)$ be the vector of budgets the ad allocator extracts in displaying ad $i$ for query type $j$ during $[0,T]$. Therefore, each query rewriting allocation strategy determines the tuples $\left( j, Y_j, \Bv^j \right)$ where $j \in \mathcal{Q}$ is a query type and $Y_j \subseteq \mathcal{R}$ is the set of rewrites for query type $j$.


We next provide an alternative formulation of query rewriting problem. To simplify the exposition, we introduce this alternative formulation for $d=1$. All the results continue to hold for any general $d > 1$ as we show in Appendix \ref{App:General}. 

\begin{definition}[Partial Configuration and Compact Allocation Strategy]\label{def:Partial}
\textup{
We call the tuple $\left( j, Y_j, \Bv^j \right)$ a \emph{partial configuration}, where $j \in \mathcal{Q}$ is a query type, $Y_j \subseteq \mathcal{R}$ such that $|Y_j| \le k$ is the set of rewrites for query type $j$, and $\Bv^j = \left( B_{1j},
\cdots, B_{mj} \right)$ is a vector of budgets in which $B_{ij}$ is the maximum budget that we allow the ad allocator to extract in displaying ad $i$ for query type $j$. We let $\mathcal{S}^{p}$ be the set of all partial configurations. We call a discrete sequence $\tilde{H}$ from the elements of $\mathcal{S}^{p}$ a \emph{compact allocation strategy}, which is of the form 
\begin{align*}
\tilde{H}= \left(	\left( j_1, Y_{j_1}, \Bv^{j_1} \right), \cdots, \left( j_n, Y_{j_n}, \Bv^{j_n}\right) \right),
\end{align*}
where $j_1, \dots, j_n$ is a permutation of $1, \dots, n$. 
$\mathds{H}^D(\mathcal{S}^{p})$ denotes the set of all compact allocation strategies. 
}
\end{definition}

 For any compact allocation strategy $\tilde{H}$, its revenue function denoted by $\tilde{u}(\tilde{H})$ is the revenue collected by sequentially running the partial configurations specified by sequence $\tilde{H}$ and extract the optimal revenue given the budget constraints. The budget constraints are imposed by both the remaining budget of the advertisers and the budget vector of the current partial configuration. We next formally define the revenue function of a compact allocation strategy.

 

\begin{definition}\label{def:Utility}[Revenue Function of Compact Allocation Strategy]
For a given compact allocation strategy $\tilde{H}$ we recursively define its revenue function. We let $\tilde{u}(\emptyset)$ be zero and initialize the current budget of advertiser $i$ denoted by $\tilde{B}_i$ as the original budget of advertiser $i$, i.e., $B_i$. Suppose $( j, Y_{j}, \Bv^j)$ is the current element of $\tilde{H}$ (with the order specified by $\tilde{H}$). The budget limit of advertiser $i$ is the minimum of the current budget $\tilde{B}_i$ and $B_{ij}$. For $T q_j$ queries of type $j$, considering the remaining budget limits and the given rewrite set $Y_j$, we greedily select the optimal query rewriting configuration. That is for the first query of type $j$ we use configuration $s$ such that 
\begin{align*}
s(j) \in \argmax_{i \in \bigcup_{r \in Y_j} W_r} p_{ij},
\end{align*}
where the advertisers without budget are deleted from the set $\bigcup_{r \in Y_j} W_r$. We then use this configuration until we meet the budget constraint of advertiser $s(j)$. We then update the configuration and switch to the ad with the second highest payment. We continue this approach for all $T q_j$ queries of type $j$. We also assume whenever an advertiser runs out of budget (i.e., its budget is less than the payment of the ad specified by the configuration in use) the ad allocator shows the ad for a fraction of time and charges the advertiser for that fraction. We add the collected payments to the current $\tilde{u}(\tilde{H})$ and update $\tilde{B}_i$ by subtracting the consumed budget of advertiser $i$ in serving query type $j$. We then proceed to the partial configuration in the sequence $\tilde{H}$ and the revenue function $\tilde{u}(\tilde{H})$ is the revenue obtained at the end of this procedure. 
\end{definition}
Note that any compact allocation strategy defines a query rewriting allocation strategy in which for any arriving query the ad allocator uses the configuration specified in computing revenue function of compact allocation strategy $\tilde{u}$ (Definition \ref{def:Utility}). Moreover, for any query rewriting allocation strategy, the revenue obtained from its corresponding compact allocation strategy (where the query types are ordered in an arbitrary order in the sequence) is exactly the same as the revenue from the query rewriting allocation strategy itself. This is because for all query types the ad allocator shows each ad with the same frequency in both allocation strategies (potentially in a different order though).\footnote{Note that since the budget consumed by all query types in the query rewriting allocation strategy is at most the advertiser's budget, in the compact allocation strategy any order of query types obtains the same revenue as the original query allocation strategy. However, for a general compact allocation strategy, different orders of the query types in the discrete sequence generates different amount of revenue (when the advertiser budgets become binding).} 

Using this alternative formulation, the query rewriting problem \eqref{eq:QR_1} is equivalent to find $Y_1, \dots, Y_n$ together with a compact allocation strategy that maximizes $\tilde{u}$, i.e., 
\begin{align}\label{Eq:QR_2}
\max_{Y_{1}, \dots, Y_{n}, \tilde{H} \in \mathds{H}(\mathcal{S}^p)}  & ~\tilde{u}(\tilde{H})\\
\text{ s.t. } & ~ |Y_{j}| \le k,~ \forall j \in \mathcal{Q}. \nonumber
\end{align}
In the next section, we provide an algorithm for maximizing discrete sequence functions and then use it to find the optimal compact allocation strategy (i.e., to solve \eqref{Eq:QR_2}). 
%
%



\section{Discrete Sequence-Submodular Functions and their Maximization}\label{sec:Dsicrete}
In this section, we define our framework for maximizing discrete sequence submodular functions and then apply it to query rewriting problem. 
\subsection{Discrete Sequences: Definition and Operations}\label{sec:discDef}

We let $\mathcal{S}$ be a finite set of actions. Any sequence $A=(s_1, \cdots, s_k)$ where $k \in \mathbb{N}\cup \{0\}$
and $s_i \in \mathcal{S}$ is called a \emph{discrete sequence} of elements of $\mathcal{S}$ (with convention, for $k=0$ we have an empty sequence). 
The length of a discrete sequence $A= (s_1, \cdots, s_k)$ denoted by $|A|$ is equal to $k$. 
We again denote the set of all finite discrete sequences of $\mathcal{S}$ by $\mathds{H}^D(\mathcal{S})$, formally defined as 
\begin{align*}
    \mathds{H}^D(\mathcal{S}) = \left\{A = (s_1, \cdots, s_k) ~|~ k \in \mathbb{N}\cup\{0\} , s_i \in \mathcal{S} \right\}. 
\end{align*}

Equivalency, concatenation and domination are similar to the continuous setting described in Section \ref{sec:model}. The refinement of a discrete sequence $A = (s_1, \cdots, s_k)$ in the set $\{x, x+1, \dots, y\}$ denoted by $A_{[x,y]}$ is $A_{[x,y]}=(s_{\max\{x, 1\}}, \cdots, s_{\min\{y,k\}})$. To be consistent with the continuous setting, we represent the set $\{x, x+1, \dots, y\}$ by $[x,y]$.

\subsection{Submodular Non-decreasing Discrete Sequence Functions}
    \label{sec:conds}
In this subsection, we define the class of submodular non-decreasing discrete sequence functions. 
Given a finite set $\mathcal{S}$, any function $u : \mathds{H}^D(\mathcal{S}) \rightarrow \mathbb{R}$ is a discrete sequence function. 

Similar to the continuous setting, a discrete sequence function $u$ is \emph{sequence-non-decreasing} if 
\begin{align}
    \label{eq:mono_2}%
    & u(A) \le u(B), \quad \forall A, B \in \mathds{H}^D(\mathcal{S}) \text{ such that } A \prec B,  \nonumber \\
    & u(\emptyset) = 0.
\end{align}
Also, a discrete sequence function $u$ is \emph{sequence-submodular} if
\begin{equation}\label{eq:ncomp}
u(C|A) \ge u(C|B), \quad  \forall A, B, C \in \mathds{H}^D(\mathcal{S}) \text{ such that } A \prec B, 
\end{equation}
where $u(B|A)$ is the \emph{marginal value} of the sequence function defined as 
\begin{align*}
	u(B|A) = u(A \bot B)-u(A), \quad  \forall A,B \in \mathds{H}^D(\mathcal{S}). 
\end{align*}

In the next section, we provide a greedy algorithm to maximize any discrete sequence function satisfying the following assumption.
\begin{assumption}\label{Assump:SS+ND}
The discrete sequence function $u$ is sequence-non-decreasing and sequence-submodular. 
\end{assumption}


%
%
%
%
%

\subsection{Greedy Algorithm for Maximizing Discrete Sequence Functions} \label{sec:MaximizationDisc}

For a given $\mathcal{S}$, a discrete sequence function $u : \mathds{H}^D(\mathcal{S}) \rightarrow \mathbb{R}$, and a length constraint $T \in
\mathbb{N}$, the objective is to find a sequence $H \in \mathds{H}^D(\mathcal{S})$ that maximizes $u$ subject to $|H| \le T$, i.e., 
\begin{align}\label{eq:DiscreteCase}
\max_{H \in \mathds{H}^D(\mathcal{S})} & u(H)  \nonumber \\
\text{ s.t. }& |H| \le T.
\end{align}

Our key result presented next establishes the performance guarantee of a greedy algorithm for maximizing discrete sequence functions.

\begin{theorem}
    \label{thm:dis}
Suppose Assumption \ref{Assump:SS+ND} holds for a discrete sequence function $u$. For any $\alpha \in [0, 1]$ and a sequence $H=(s_1, \cdots, s_T)$ in $\mathds{H}^D(\mathcal{S})$, if for all $i \in \{1, \cdots, T\}$ we have 
\begin{equation}
    \label{eq:thm:dis}
     u\left( s_i|H_{[1,i-1]} \right) \ge \alpha \max_{s \in \mathcal{S}} u(s|H_{[1,i-1]}),
\end{equation}
then
\begin{equation}
    \frac{u\left( H \right)}{u\left( O \right)} \ge 1-\frac{1}{e^\alpha},
\end{equation}
where $O \in \mathds{H}^D(\mathcal{S})$ denotes the optimal solution of problem \eqref{eq:DiscreteCase}.\footnote{The optimal solution exists because there are finitely many sequences of length at most $T$. In particular, there are no more than $(|\mathcal{S}|+1)^{T}$ many sequences $H \in \mathds{H}^D(\mathcal{S})$ with $|H| \le T$, where $|\mathcal{S}|$ denotes the cardinality of set $\mathcal{S}$.} 
\end{theorem}

%
%
%
%
%

Theorem \ref{thm:dis} shows that if the elements of the sequence $H = (s_1, \cdots, s_T)$ are chosen sequentially such that for each $i$, $u\left( s_i|H_{[1,i-1]} \right)$ is at least $\alpha$ times its optimal local maximum, then the resulting sequence yields $1-{1}/{e^\alpha}$ of the optimal sequence.  

We next outline the key idea of this result for $\alpha=1$ (the complete proof is given in the Appendix). First, using sequence-submodularity, we show that the marginal increase in function value with the greedy choice is as large as the time average marginal increase by concatenating any other sequence. Formally, for all $B \in \mathds{H}^D(\mathcal{S})$ and $t \in [0, T]$ we show  
\begin{align*}
\max_{s \in \mathcal{S}} u \left( s|H_{[0,t]} \right) \ge \frac{1}{|B|} u(B | H_{[0,t]} ).
\end{align*}
Substituting the optimal sequence, i.e., $O$, for $B$ and then using non-decreasing property, we show that this marginal increase is as large as the time average difference between function value of the optimal solution and function value of the current sequence. Formally, we have  
\begin{align*}
\max_{s \in \mathcal{S}} u\left( s|H_{[0,t]} \right) \ge \frac{1}{T} \left( u(O ) - u(H_{[0,t]} ) \right).
\end{align*}
This provides a recursive relation between the function value of the greedy choice and the optimal choice. Using this recursive relation we then establish that the utility of sequence $H$ is at least $1- 1/e$ fraction of the function value of the optimal solution $O$.

Motivated by Theorem \ref{thm:dis}, Algorithm \ref{alg:discrete} presents our greedy algorithm for maximizing discrete sequence functions satisfying Assumption \ref{Assump:SS+ND}. 

\begin{algorithm}
    \caption{Greedy algorithm for discrete setting \label{alg:discrete}}
    $H \leftarrow \emptyset$ \;

    \For{$i=1$ \KwTo $T$}{
        find $s_i$ such that $u(s_i | H) \ge \alpha \max_{s \in \mathcal{S}} u(s | H)$ \nllabel{alg:discrete:max}\;
        $H \leftarrow H \bot s_i$ \;
    }
\end{algorithm}

Algorithm \ref{alg:discrete} starts with an empty sequence $H$ (the initialization is $H=\emptyset$) and at each step adds one element to the sequence. 
For instance with $\alpha=1$, at step $i$ the algorithm finds $s_i$ that generates the highest increase in the value of $u$ when concatenated to the end of
the current sequence (i.e., finds $s_i$ that maximizes $u(s_i|H)$). Also note that in Algorithm \ref{alg:discrete}, at step $i$ the problem of finding $s_i$ that maximizes $u(s_i
| H)$ may be computationally hard. Theorem \ref{thm:dis} states that even if $u(s_i|H)$ is $\alpha$ times the local maximum, Algorithm \ref{alg:discrete} still provides a good approximation of the optimal solution, namely $1- {1}/{e^{\alpha}}$ approximation. For instance, in Subsection \ref{sec:queryApp}, we show that for query rewriting, the local problem at each step can efficiently be solved with $\alpha= 1-{1}/{e}$.

The following is an immediate corollary of Theorem \ref{thm:dis} and formally states the performance of Algorithm
\ref{alg:discrete}.


\begin{corollary}
    \label{thm:dis:fin}
Suppose Assumption \ref{Assump:SS+ND} holds for a discrete sequence function $u$. Algorithm
\ref{alg:discrete} generates a discrete sequence with value at least $1-{1}/{e^\alpha}$ of the optimal solution.
\end{corollary}

We next show how to apply Theorem \ref{thm:dis} and Algorithm \ref{alg:discrete} to query rewriting problem.

\subsection{Application to Query Rewriting}\label{sec:queryApp}
We first show that the revenue function of compact allocation strategy is non-decreasing and sequence-submodular. We then show that the local optimization problems of the greedy algorithm can be solved with $\alpha= 1- 1/e$. Finally, we use our analysis of Subsection \ref{sec:MaximizationDisc}, which finds a greedy algorithm  for the compact allocation strategy (and equivalently, query rewriting problem) that achieves $1- {1}/{e^{1-\frac{1}{e}}}$ of the optimal revenue.

We start by showing the function $\tilde{u}:\mathds{H}^D(\mathcal{S}^{p}) \to \mathbb{R}$ satisfies Assumption \ref{Assump:SS+ND}. 
\begin{lemma}\label{Lem:QuerySSGreedy}
The revenue function of compact allocation strategies, i.e., $\tilde{u}:\mathds{H}^D(\mathcal{S}^{p}) \to \mathbb{R}$ is sequence-non-decreasing and sequence-submodular.
\end{lemma}
The proof of this lemma is similar to that of Lemma \ref{thm:onlineAd} and is given in the appendix. 

We next describe a greedy algorithm for maximizing $\tilde{u}(\cdot)$ and then use Theorem \ref{thm:dis} to establish its performance.
The greedy algorithm works as follows. At each step of the algorithm with the current allocation $\tilde{H}$ and current budget vector $\Bv$, for any remaining query type such as $j$ we greedily find the optimal set of rewrites $Y_j$ with cardinality constraint $|Y_j| \le k$. We then greedily select the tuple $(j, Y_{j}, \Bv^j)$ whose addition to the current strategy increases the revenue the most. Finally, we update the advertisers' budgets by subtracting $\Bv^j$ and update $H$ by appending $(j, Y_{j}, \Bv^j)$ to it. The complete algorithm is described in Algorithm \ref{alg:rew}.
\begin{algorithm}[t]
    \caption{Query Rewriting Algorithm \label{alg:rew}}
    $\tilde{H} \leftarrow \emptyset$ \;
    $\Bv \leftarrow (B_1, \cdots, B_m)$ \;
    \While{$\mathcal{Q} \neq \emptyset$}{
        
        \For{$j \in \mathcal{Q}$}{
            $Y_j \leftarrow \emptyset$ \;
            \tcp{find the best $k$ rewrites greedily}
            \For{$w=1, \cdots, k$}{
                $r \leftarrow  \argmax_{r' \in R \setminus Y_j} \tilde{u}\left( \left(j, Y_j\cup \{r'\}, \Bv \right)| \tilde{H} \right)$ \;
                $Y_j \leftarrow Y_j \cup \{r\}$ \;
            }
            
        }
        
        \tcp{find the best partial configuration to append}
            $j \leftarrow \argmax_{j' \in \mathcal{Q}}  \tilde{u}\left( \left(j', Y_{j'}, \Bv \right)| \tilde{H} \right)$
        
        Define $\Bv^{j}$ as the amount of budget used by $(j, Y_j, \Bv)$ when appended to $\tilde{H}$ \;
        $\tilde{H} \leftarrow \tilde{H} \bot (j, Y_{j}, \Bv^{j})$ \;
        $\mathcal{Q} \leftarrow \mathcal{Q} \setminus \{j\}$ \;
        $\Bv \leftarrow \Bv-\Bv^{j}$ \;
    }
\end{algorithm}

Our next lemma establishes the performance of Algorithm \ref{alg:rew}. 

\begin{lemma}\label{Lem:mooaw}
Algorithm \ref{alg:rew} finds a compact allocation strategy whose revenue function is at least $1- {1}/{e^{1- \frac{1}{e}}} $ of the optimal compact allocation strategy.
\end{lemma}
This lemma follows from applying Algorithm \ref{alg:discrete} to maximize $\tilde{u}(\cdot)$ and then using Theorem \ref{thm:dis} with $\alpha= 1-{1}/{e}$. In particular, in the local optimization step of Algorithm \ref{alg:discrete}, we greedily find a set of rewrites subject to a cardinality constraint. We show that the greedy approach solves the local optimization problem with $\alpha=1- 1/e$. The result then follows from using Lemma \ref{Lem:QuerySSGreedy} and Theorem \ref{thm:dis}. The complete proof is given in the appendix.

The output of Algorithm \ref{alg:rew} naturally defines a query rewriting allocation strategy as follows. For all $j \in \mathcal{Q}$, we let $Y_j$ be the set of rewrites found by Algorithm \ref{alg:rew}. For arriving queries of type $j$, we use the query configurations specified in computing $\tilde{u}$ of the output of Algorithm \ref{alg:rew}. 
We next find the performance guarantee of this query rewriting allocation strategy. 
\begin{theorem}\label{pro:Equivalent}
The query rewriting allocation strategy defined based on the output of Algorithm \ref{alg:rew} achieves 
\begin{align*}
\left( 1- \frac{1}{e^{1- \frac{1}{e}}}\right) - \left( \frac{\min_{j \in \mathcal{Q}, i \in \mathcal{A}} p_{ij}}{ \min_{i \in \mathcal{A}} B_i} \right)
\end{align*}
of the optimal query rewriting allocation strategy. 
\end{theorem}
Theorem \ref{pro:Equivalent} follows from Lemmas \ref{Lem:QuerySSGreedy} and \ref{Lem:mooaw} and establishes the performance guarantee of our algorithm which holds for any bid to budget ratio. We next make a few remarks regarding Algorithm \ref{alg:rew} and its corresponding query rewriting allocation strategy. First, note that this algorithm requires the knowledge of the distribution $\mathbf{q}$ (same as the algorithm of \cite{MCKW08}). This is because in defining the function $\tilde{u}(\cdot)$ we need to know the number of queries of type $j$ in the time interval $[0,T]$. The order of incoming queries, however, is unknown and random. Second, using Theorem \ref{pro:Equivalent} for small bid to budget ratio, the query rewriting allocation strategy defined based on the output of Algorithm \ref{alg:rew} achieves $1- \frac{1}{e^{1-\frac{1}{e}}} \approx 0.47$. This outperforms the existing algorithm with approximation factor ${1}/{4}$ given in \cite{MCKW08}.

\section{Conclusion}\label{sec:Conclusion}
Motivated by applications in online advertising, we develop a framework to maximize functions defined over sequences. In particular, we extend the notion of submodularity and monotonicity for functions that are defined over sets to functions that are defined over sequences (both continuous and discrete). We then show that if a sequence function is sequence-submodular and non-decreasing (and differentiable in the case of continuous sequences), then a greedy
algorithm that solves the local optimization problems with factor $\alpha$ achieves $1 - {1}/{e}^{\alpha}$ of the maximum subject to a length
constraint.

Our framework provides a simple and yet powerful method for solving a broad class of maximization problems where the objective is defined over sequences. We demonstrated the applicability of our framework by considering two applications in online advertising problems. In particular, we showed that online ad allocation problem can be formulated as maximizing a sequence submodular function. We then used our results and algorithms to establish that assuming the bid to budget ratio is very small, an online greedy approach achieves $1- {1}/{e}$
of the optimal revenue. We then considered query rewriting problem in search advertising. For this problem, we defined an offline problem whose objective function is the same as online query rewriting problem. We then showed that a greedy algorithm achieves $1- {1}/{e^{1-\frac{1}{e}}}$ of the optimal solution of this offline problem. Finally, we used this offline solution to find an online algorithm for query rewriting problem which outperforms the existing algorithm in the literature. Avenues for future research include the study of other problems that can be formulated as maximizing a sequence-submodular function and then using our algorithm to establish performance guarantee of greedy algorithms for solving them.


\section{Appendix}
\subsection{Proofs of Section \ref{sec:model}}
\subsubsection*{Proof of Theorem \ref{thm:con}}
We first present some lemmas that we use in proving this theorem. 

\begin{lemma}
    \label{cor:con:udots}
For any $A = \left( (s_1, \Delta t_1), \cdots \Br, (s_k, \Delta t_k) \right)$ in $\mathds{H}^C(\mathcal{S})$, we have 

\begin{align}
    u\left( (s,\delta)|A \right) & = \int_0^\delta \dot{u}_s(x|A) dx \label{eq:con:int} \\
    u\left( (s,\delta_2)|A\bot(s,\delta_1) \right) & = \int_{\delta_1}^{\delta_2} \dot{u}_s(x|A) dx \label{eq:con:int2} \\
    u(A) & = \sum_{i=1}^k \int_0^{\Delta t_i} \dot{u}_{s_i}(x|A^{i-1}) dx, \label{eq:con:sum}
\end{align}
where $A^i = \left( (s_1,\Delta t_1),
\cdots, (s_i, \Delta t_i) \right)$. 
\end{lemma}

\begin{proof}

Eqs. \eqref{eq:con:int} and \eqref{eq:con:int2} directly follow from Eq. \eqref{eq:con:udot}, and Eq. \eqref{eq:con:sum} directly follows from the definition of marginal values. 
\end{proof}
The next two lemmas show diminishing return property, i.e., $\dot{u}_s\left( \delta|A \right)$
is decreasing in both $\delta$ and $A$. More specifically, using Assumption \ref{Assump:SS+ND+Diff} we show for any $A$, $\dot{u}_s\left( \delta|A \right)$ is decreasing as a function of $\delta$. Moreover, if $A \prec B$, then $\dot{u}_s\left( \delta|A \right) \ge \dot{u}_s\left( \delta|B \right)$ (except at finitely many points). 

\begin{lemma}
    \label{lem:udot:ncomp}
Suppose Assumption \ref{Assump:SS+ND+Diff} holds for continuous sequence function $u$. For any $A,B \in \mathds{H}^C(\mathcal{S})$ such that $A \prec B$ and any $s \in \mathcal{S}$, we have $\dot{u}_s(\delta|A) \ge
\dot{u}_s(\delta|B)$ for all $\delta \in \mathbb{R}^+\cup \{0\}$ except at a finite number of points.
\end{lemma}
\begin{proof}
We prove this lemma by contradiction. Suppose there are $A,B \in \mathds{H}^C(\mathcal{S})$ such that $A \prec B$ and $s \in \mathcal{S}$ and $\delta
\in \mathbb{R}^+$ for which $\dot{u}_s(\delta|A) < \dot{u}_s(\delta|B)$. If either $\dot{u}_s(\delta|A)$ or
$\dot{u}_s(\delta|B)$ is non-continuous at $\delta$ then this is one of the finite number of points that are exceptions
in the statement of Lemma \ref{lem:udot:ncomp}. Otherwise, since they are both continuous at $\delta$ there should be a small neighborhood
around $\delta$ in which $\dot{u}_s(\delta|B)$ is greater than $\dot{u}_s(\delta|A)$. More formally, there exists $ \epsilon \in \mathbb{R}^+$ such that for all $x \in [\delta-\epsilon,\delta+\epsilon]$, we have 

\begin{equation}
    \label{eq:udot:ncomp}
\dot{u}_s(x|A) < \dot{u}_s(x|B). 
\end{equation}

We next show that Eq. \eqref{eq:udot:ncomp} can never happen. Using Lemma \ref{cor:con:udots} and in particular \eqref{eq:con:int2}, we have 

\begin{align*}
    u((s,\epsilon)|A \bot (s,\delta-\epsilon)) & = \int_{\delta-\epsilon}^\delta \dot{u}_s(x|A),  \\
    u((s,\epsilon)|B \bot (s,\delta-\epsilon)) & =\int_{\delta-\epsilon}^\delta \dot{u}_s(x|B).
\end{align*}

These equalities together with Eq. \eqref{eq:udot:ncomp}, leads to 

\begin{align}\label{eq:udot:ncomp:contradiction}
    u\left( (s,\epsilon)|A \bot (s,\delta-\epsilon) \right) & < u\left( (s,\epsilon)|B \bot (s,\delta-\epsilon) \right).
\end{align}

Since $A \bot (s,\delta-\epsilon) \prec B \bot (s,\delta-\epsilon)$, Assumption \ref{Assump:SS+ND+Diff} and in particular sequence submodularity of $u$ results in 
\begin{align}\label{eq:udot:ncomp:contradiction2}
    u\left( (s,\epsilon)|A \bot (s,\delta-\epsilon) \right) & \ge u\left( (s,\epsilon)|B \bot (s,\delta-\epsilon) \right).
\end{align}
Eqs. \eqref{eq:udot:ncomp:contradiction} and \eqref{eq:udot:ncomp:contradiction2} contradict each other, showing that our assumption of $\dot{u}_s(\delta|A) < \dot{u}_s(\delta|B)$ does not hold. This completes the proof. 

\end{proof}
\begin{lemma}
    \label{cor:con:dec}
Suppose Assumption \ref{Assump:SS+ND+Diff} holds for continuous sequence function $u$. For any $A \in \mathds{H}^C(\mathcal{S})$, and any $\delta \in [0, \infty)$, $\dot{u}_s(\delta|A)$ is a monotonically
non-increasing function in $\delta$. That is for $\delta_1 < \delta_2$, we have $\dot{u}_s(\delta_1|A) \ge
\dot{u}_s(\delta_2|A)$.
\end{lemma}
\begin{proof}
Using Eq. \eqref{eq:con:udot} we have
\begin{align}
\label{eq:boj}\dot{u}_s(\delta_2|A)&=\frac{d}{d x}u((s, x)|A) \Bigr |_{x=\delta_2} =\frac{d}{d x}u((s, x)|A\bot(s,\delta_2-\delta_1) \Bigr |_{x=\delta_1}=\dot{u}_s(\delta_1|A\bot (s,\delta_2-\delta_1)).
\end{align}
Since $A\prec A\bot (s,\delta_2-\delta_1)$, combining Lemma \ref{lem:udot:ncomp} and Eq. \eqref{eq:boj} implies that $\dot{u}_s(\delta_2|A)<\dot{u}_s(\delta_1|A)$, completing the proof.
\end{proof}

\begin{lemma}
    \label{lem:dv}
Suppose Assumption \ref{Assump:SS+ND+Diff} holds for continuous sequence function $u$. For any $A,B \in \mathds{H}^C(\mathcal{S})$ there exists $s \in \mathcal{S}$ such that $\dot{u}_s(0|A) \ge \frac{1}{|B|}u(B|A)$
\end{lemma}

\begin{proof}

Letting $B=((s_1, \Delta t_1), \cdots, (s_k, \Delta t_k))$ and $B^i = ((s_1,\Delta t_1), \cdots, (s_i, \Delta
t_i))$, and using the definition of $u$ and Lemma \ref{cor:con:udots} we obtain 

\begin{equation}
    \label{eq:con:dv:sum}
    u(B | A) = \sum_{i=1}^k \int_0^{\Delta t_i} \dot{u}_{s_i}\left( x|A \bot B^{i-1} \right) dx.
\end{equation}
We argue that there should be some $1 \le i \le k$ for which there exists some $\delta \in [0, \Delta t_i)$ such that
$\dot{u}_{s_i}\left( \delta |A \bot B^{i-1} \right) \ge \frac{1}{|B|}u(B|A)$. Otherwise, the term inside the integral on
the right hand side of Eq. \eqref{eq:con:dv:sum} is always less than $\frac{1}{|B|}u(B|A)$ which means that the sum of the
integrals is less than $u(B|A)$, which contradicts Eq. \eqref{eq:con:dv:sum}. Suppose for $i'$ and $\delta'$
we have 
\begin{align}\label{eq:con:dv:B}
    \dot{u}_{s_{i'}}\left( \delta' |A \bot B^{i'-1} \right) & \ge \frac{1}{|B|}u(B|A).
\end{align}
Using Lemma \ref{lem:udot:ncomp} in Eq. \eqref{eq:con:dv:B} leads to 
\begin{align}\label{eq:con:dv:delta}
    \dot{u}_{s_{i'}}(\delta' |A) & \ge \frac{1}{|B|}u(B|A).
\end{align}
Finally, invoking Lemma \ref{cor:con:dec} in Eq. \eqref{eq:con:dv:delta}, yields
\begin{align*}
    \dot{u}_{s_{i'}}(0|A) & \ge \frac{1}{|B|}u(B|A),
\end{align*}
which completes the proof. 

\end{proof}

We next proceed with the proof of theorem. Consider a sequence $H$ and $\alpha$ for which
Eq. \eqref{eq:thm:con} holds. 
Using Lemma \ref{lem:dv}, for all  $t \in [0, T)$, there exists $s \in \mathcal{S}$ such that 
\begin{align}\label{eq:thm:con:dv}
 \dot{u}_s\left( 0|H_{[0,t)} \right) \ge \frac{1}{|O|} u\left( O|H_{[0,t)} \right).
\end{align}
Using Eq. \eqref{eq:thm:con} in Eq. \eqref{eq:thm:con:dv}, leads to 
\begin{align}\label{eq:thm:con:ncomp}
\frac{d}{dt}u\left( H_{[0,t)} \right) \ge \frac{\alpha}{T} u\left( O|H_{[0,t)} \right), \quad \forall t \in [0, T). 
\end{align}
Using the definition of marginal values, we can rewrite Eq. \eqref{eq:thm:con:ncomp} as the following differential equation. 
\begin{align}\label{eq:thm:con:de0}
\frac{d}{dt}u(H_{[0,t)}) \ge \frac{\alpha}{T} \left( u(O \bot H_{[0,t)})-u(H_{[0,t)}) \right), \quad \forall t \in [0, T).
\end{align}
Using Assumption \ref{Assump:SS+ND+Diff} in Eq. \eqref{eq:thm:con:de0} results in 
\begin{align*} 
\frac{d}{dt}u(H_{[0,t)}) \ge \frac{\alpha}{T} \left( u(O)-u(H_{[0,t)}) \right), \quad \forall t \in [0, T), 
\end{align*}
or equivalently 
\begin{align}\label{eq:thm:con:de2} 
    u(H_{[0,t)})+\frac{T}{\alpha} \frac{d}{dt}u(H_{[0,t)})  \ge u(O) , \quad \forall t \in [0, T). 
\end{align}
We can rewrite Eq. \eqref{eq:thm:con:de2} as 
\begin{align*}
    \frac{d}{dt}\left(\frac{T}{\alpha} e^{\frac{\alpha}{T}t}u(H_{[0,t)})\right) \ge \frac{T}{\alpha} e^{\frac{\alpha}{T}t} u(O), \quad \forall t \in [0, T). 
\end{align*}
Therefore, for any $x \in (0, T]$, we have 
\begin{align*}
    \int_0^x \frac{d}{dt}\left(\frac{T}{\alpha} e^{\frac{\alpha}{T}t}u(H_{[0,t)})\right) dt \ge \int_0^x e^{\frac{\alpha}{T}t} u(O) dt.
\end{align*}
Computing the integral of both sides, leads to 
\begin{align*}
    \frac{T}{\alpha} e^{\frac{\alpha}{T}x}u\left( H_{[0,x)} \right) & \ge \frac{T}{\alpha} \left( e^{\frac{\alpha}{T}x}-1 \right) u(O), \quad \forall x \in (0, T],
\end{align*}
which in equivalent to 
\begin{align}\label{eq:thm:con:fin0}
    u(H_{[0,x)}) & \ge \left(1-\frac{1}{e^{\frac{\alpha}{T}x}} \right) u(O), \quad \forall x \in (0, T]. 
\end{align}
Finally, letting $x=T$ in Eq. \eqref{eq:thm:con:fin0} yields
\begin{align*}
    u(H) & \ge \left( 1-\frac{1}{e^\alpha} \right) u(O),
\end{align*}
 which completes the proof. 
\subsubsection*{Proof of Lemma \ref{thm:onlineAd}}
We prove this lemma in three steps. 
\newline \textbf{Step 1:} In this step, we show that the revenue function of online ad allocation problem satisfies monotonicity.
In particular, consider the allocation strategies $A,B \in \mathds{H}(\mathcal{S})$ and assume that $A \prec B$. 
We argue that the revenue
extracted from each ad after running sequence $B$ is at least as much as the revenue extracted from each ad after running sequence $A$. We partition the ads into two categories:
\begin{enumerate}
\item
Ads that have no budget left after running sequence $B$. Note that in this variation of online ad allocation problem, when an advertiser such as $i$ runs out of budget its total budget $B_i$ is used.    
\item
Ads that still have budget after running sequence $B$.
\end{enumerate}
For the ads in the first category, sequence $B$ has extracted the maximum possible budget from the ad. Therefore, for this set of ads our claim holds. For the ads that belong to the second category, we know that they still have budget available. We consider an ad
$i$ that belongs to this category and show that the revenue extracted by $B$ from this ad is at least as
much as the revenue extracted by $A$. Consider the configuration $s \in \mathcal{S}$ that is being used in $B$ for a total time of $\Delta t$. For all queries of
type $j$ that arrive during this time and any ad $i$ that is allocated to them by configuration $s$, we know
that the revenue extracted from budget of ad $i$ by those queries is $\Delta t \ p_{ij}$ because ad $i$ never ran
out of budget. Since $A \prec B$, configuration $s$ is either not present in $A$ or was used in $A$ for a duration of no more than $B$. Thus, the total revenue extracted from ad $i$ in $A$ is no more than the revenue extracted
from ad $i$ in $B$. Since for both categories of ads, the expected revenue extracted by $B$ from the ads are higher than or equal to the
revenue extracted by $A$ from the ads, we conclude that the sequence-non-decreasing property holds.
\newline \textbf{Step 2:} In this step, we show that the revenue of the online ad allocation problem satisfies sequence-submodularity. Consider the allocation strategies $A,B,C \in \mathds{H}(\mathcal{S})$ and assume that $A \prec B$. First of all, based on
Step 1, we know that the remaining budget of each ad after $A$ is greater than or equal to
its remaining budget after $B$. Moreover, the contribution of each ad to $\tilde{u}(C|B)$ or
$\tilde{u}(C|A)$ is equal to the difference in its budget before and after using the sequence $C$. Now, consider using the
allocation strategy $A$ first, followed by strategy $C$. Again, we partition the ads into two categories:

\begin{enumerate}
\item Ads that have exhausted all of their budget after running $A\bot C$.

\item Ads that still have budget after running $A \bot C$.
\end{enumerate}
The contribution of the ads in the first category to $\tilde{u}(C|B)$ is no more than their contribution to $\tilde{u}(C|A)$
because as was shown in Step 1, they had equal or more remaining budget after using $A$ than after using $B$ and they have
contributed all of their remaining budget to $\tilde{u}(C|A)$. Now consider the ads that belong to the second category. Using the same reasoning as we did for the proof of
Step 1, we conclude that $C$ has extracted revenue from those ads at full rate since
they did not run out of budget. Thus, their contribution to $\tilde{u}(C|A)$ is larger than (or equal to) its contribution to $\tilde{u}(C|B)$. 
\newline \textbf{Step 3:} In this step, we show that the revenue of the online ad allocation problem satisfies differentiability. Note that the derivative of the revenue function
is a step function that only changes its value when there is change of a configuration in the sequence or when some ad runs out of the budget. We next show that, without loss of generality, we can only consider sequences with $k \le |\mathcal{S}| m$ where the number of configurations is bounded by $|\mathcal{S}| \le m^{dn}$ (note that the number of configurations that our greedy algorithm finds is much smaller than this and is in fact bounded by the number of ads $m$). Let $O$ be the optimal sequence and let $t$ and $t'$ be two consecutive times at which an ad runs out of budget. We next show that the number of configurations used in this interval is bounded by $|\mathcal{S}|$, showing the overall number of configurations is bounded by $|\mathcal{S}| m$. In the time interval $(t, t')$, if a configuration $s$ is used multiple times, then we can move all those times together to form an interval during which configuration $s$ is used. This does not change the revenue function because no ad has run out of budget in the interval $(t, t')$. Thus, the revenue function is differentiable and its derivative is continuous except at a finite number of points.

\subsubsection*{Proof of Lemma \ref{Lem:jaDd_1}}
We let $\tilde{O}$ and $\tilde{H}$ denote the optimal allocation and the allocation strategy obtained by running greedy algorithm for the variation of online ad allocation problem. We also let $O$ and $H$ denote the optimal allocation and the allocation strategy obtained by running greedy algorithm for the original online ad allocation problem. 

Using Proposition \ref{pro:OnlineAdAlloc} and Theorem \ref{thm:con}, we have 
\begin{align}\label{eq:tomp_1}
\tilde{u}(\tilde{H}) \ge \left(1- \frac{1}{e} \right) \tilde{u} (\tilde{O}).
\end{align}
The sequence $\tilde{H}$ (obtained from using the greedy algorithm for the relaxed variation) is identical to the sequence $H$ except when one of the ads runs out of the budget. Therefore, we have  
\begin{align}\label{eq:tomp_2}
u(H) \ge \tilde{u} (\tilde{H}) - \sum_{i \in \mathcal{A},~ i\text{'s budget in } \tilde{H} \text{ is exhausted }} p_{ij}. 
\end{align}
Using Eqs. \eqref{eq:tomp_1} and \eqref{eq:tomp_2} leads to 
\begin{align}\label{eq:tomp_3}
u(H) \ge  \left(1- \frac{1}{e} \right) \tilde{u} (\tilde{O}) - \sum_{i \in \mathcal{A},~ i\text{'s budget in } \tilde{H} \text{ is exhausted }} p_{ij}.
\end{align}
We also have 
\begin{align}\label{eq:tomp_4}
\sum_{i \in \mathcal{A},~ i\text{'s budget in } \tilde{H} \text{ is exhausted }} p_{ij} & \le \left( \frac{\max_{i \in \mathcal{A}, j \in \mathcal{Q}} p_{ij}}{ \min_{i \in \mathcal{A}} B_i} \right) \sum_{i \in \mathcal{A},~ i\text{'s budget in } \tilde{H} \text{ is exhausted }} B_i  \nonumber \\
& \le  \left(\frac{\max_{i \in \mathcal{A}, j \in \mathcal{Q}} p_{ij}}{ \min_{i \in \mathcal{A}} B_i} \right) \tilde{u}(\tilde{H}) \le \left( \frac{\max_{i \in \mathcal{A}, j \in \mathcal{Q}} p_{ij}}{ \min_{i \in \mathcal{A}} B_i} \right) \tilde{u}(\tilde{O}). 
\end{align}
Invoking \eqref{eq:tomp_4} in \eqref{eq:tomp_3}, leads to 
\begin{align*}
u(H) & \ge  \left(1- \frac{1}{e} \right) \tilde{u} (\tilde{O}) - \left( \frac{\max_{i \in \mathcal{A}, j \in \mathcal{Q}} p_{ij}}{ \min_{i \in \mathcal{A}} B_i} \right) \tilde{u}(\tilde{O}) \\
& \ge \left(1- \frac{1}{e} \right) \tilde{u} (O) - \left( \frac{\max_{i \in \mathcal{A}, j \in \mathcal{Q}} p_{ij}}{ \min_{i \in \mathcal{A}} B_i} \right) \tilde{u}(O).
\end{align*}
This completes the proof. 
\subsubsection*{Proof of Lemma \ref{Lem:DerivativeOnlineAdAlloc}}
Using Eqs. \eqref{eq:con:udot} and \eqref{eq:con:udot:last} we obtain 
\begin{align*}
\dot{u}_s(0|H_{[0,t)}) & = \lim_{\delta \rightarrow 0^+} \dot{u}_s(\delta|H_{[0,t)}) = \lim_{\delta \rightarrow 0^+} \frac{d}{d x}u((s, x)|H_{[0,t)})  \Bigr |_{x=\delta}= \lim_{\delta \rightarrow 0^+} \frac{d}{d x}u(H_{[0,t)} \bot (s, x))  \Bigr |_{x=\delta}\\
& \stackrel{(a)}{=} \lim_{\delta \rightarrow 0^+} \frac{d}{d x} \left( \sum_{k=1}^{\infty} e^{-x} \frac{x^k}{k!} \mathbb{E}\left[ \text{revenue with } s \mid  k \text{ queries} \right] \right)\Bigr |_{x=\delta} \\
& = \lim_{\delta \rightarrow 0^+} \left(e^{-\delta}- \delta e^{-\delta} \right) \mathbb{E}\left[ \text{revenue with } s \mid  1 \text{ query} \right]  \\
& + \lim_{\delta \rightarrow 0^+} \sum_{k=2}^{\infty} \left( \frac{e^{-\delta} \delta^{k-1}}{(k-1)!} \frac{-e^{-\delta} \delta^{k}}{k!}\right) \mathbb{E}\left[ \text{revenue with } s \mid  k \text{ queries} \right]\\
& = \mathbb{E}\left[ \text{ revenue with } s \mid 1 \text{ query } \right] \stackrel{(b)}{=} \sum_{j \in \mathcal{Q}} q_j\sum_{i \in s(j)} p_{ij},
\end{align*}
where (a) follows from the fact that queries are arriving according to a Poisson point process with rate $1$ and therefore the number of queries in the interval $[t, t+x]$ has a Poisson distribution with rate $x$; and (b) holds because the (one arriving) query is of type $j \in \mathcal{Q}$ with probability $p_j$ in which case the revenue becomes $\sum_{i \in s(j)} p_{ij}$. 

\subsubsection*{Proof of Proposition \ref{pro:OnlineAdAlloc}}
Using Theorem \ref{thm:onlineAd}, the expected utility of the online ad allocation satisfies Assumption \ref{Assump:SS+ND+Diff}. For online ad allocation problem, using Lemma \ref{lem:dv}, we can solve the local optimization problem of Algorithm \ref{alg:cont} with $\alpha=1$. Therefore, using Theorem \ref{alg:cont} establishes Algorithm \ref{alg:cont} which is equivalent to Algorithm  \ref{alg:adallocation} for online ad allocation problem achieves $1- \frac{1}{e}$ of the optimal solution. This completes the proof.

\subsection{Proofs of Section \ref{sec:Dsicrete}}
\subsubsection*{Proof of Theorem \ref{thm:dis}}
We first show a lemma that we use in this proof. This lemma is the analogy of Lemma \ref{lem:dv} for the discrete setting.

\begin{lemma}
    \label{lem:elem}
Suppose the sequence function $u$ is sequence-submodular. For any $A,B \in \mathds{H}^D(\mathcal{S})$ there exists $s \in \mathcal{S}$ such that $u(s|A) \ge \frac{1}{|B|}u(B|A)$
\end{lemma}
\begin{proof}
Letting $B=(s_1, \cdots, s_k)$ and using the definition of sequence function $u$ we obtain 

\begin{align}\label{eq:dis:bound:p}
    u(B|A) & =   u\left( A \bot B \right)-u\left( A \right)  \nonumber \\
   & =  \sum_{j=1}^k u\left( A \bot B_{[1, j]} \right)- \sum_{j=0}^{k-1} u\left( A \bot B_{[1, j]} \right)  \nonumber \\
  & =  \sum_{j=1}^k \left( u\left( A \bot B_{[1, j]} \right) -u \left( A \bot B_{[1, j-1]}\right) \right)  \nonumber \\
  & = \sum_{j=1}^k u\left( B_{[j,j]} | A \bot B_{[1, j-1]} \right) \nonumber \\
  & =     \sum_{j=1}^k u\left( s_j | A \bot B_{[1, j-1]} \right) 
\end{align}

The sum on the right hand side of Eq. \eqref{eq:dis:bound:p} consist of $k$ terms, so there should be at least
one term which is above or equal to the average of the terms. Therefore, there exists an index $1 \le j'
\le k$ such that 
\begin{align}\label{eq:dis:bound:jpd1}
    u(s_{j'}|A\bot B_{[1, j'-1]}) & \ge \frac{1}{k} u(B|A). 
\end{align}
Using sequence-submodularity of $u$ in Eq. \eqref{eq:dis:bound:jpd1} and  because $A \prec A \bot B_{[1, j'-1]}$ we obtain  
\begin{align}\label{eq:dis:bound:jpd2}
    u(s_{j'}|A)  \ge \frac{1}{|B|} u(B|A). 
\end{align}
This completes the proof of lemma. 
\end{proof}
We now proceed with the proof of theorem. Consider a sequence $H=(s_1, \cdots, s_k)$ and $\alpha$ for which Eq. 
\eqref{eq:thm:dis} holds. Using Lemma \ref{lem:elem} we have 
\begin{align}\label{eq:dis:elem}
    u(s_i | H_{[1, i-1]}) & \ge \frac{\alpha}{k} u(O | H_{[1, i-1]})  = \frac{\alpha}{k} (u(O \bot H_{[1, i-1]})-u(H_{[1, i-1]})). 
\end{align}
Now using Assumption \ref{Assump:SS+ND}, we have $u\left(O \bot H_{[1, i-1]} \right) \ge u\left(O \right)$. This inequality together with Eq. \eqref{eq:dis:elem} leads to 
\begin{align}\label{eq:dis:elem2}
    u(s_i | H_{[1, i-1]}) & \ge  \frac{\alpha}{k} (u(O)-u(H_{[1, i-1]})).
\end{align}
Rewriting this inequality yields 
\begin{align}\label{eq:dis:elem3}
    u(H_{[1, i]})-u(H_{[1, i-1]}) & \ge \frac{\alpha}{k} (u(O)-u(H_{[1, i-1]})),
\end{align}
or equivalently 
\begin{align}\label{eq:dis:elem:fin}
    u(H_{[1, i]}) & \ge \frac{\alpha}{k}u(O) + (1-\frac{\alpha}{k}) u(H_{[1, i-1]}).
\end{align}
Using Eq. \eqref{eq:dis:elem:fin} recursively for $i=1, \dots, k$, we can bound $u\left( H_{[1, k]} \right)$ as follows
\begin{align}\label{eq:dis:final}
    u\left( H_{[1, k]} \right) & \ge \left(1-\left( 1-\frac{\alpha}{k}\right)^k\right) u(O) = \left(1-\left( \left( 1-\frac{\alpha}{k} \right)^\frac{k}{\alpha}\right)^\alpha \right) u(O).
\end{align}
Finally, invoking the inequality $\left( 1- \frac{1}{x} \right)^{x} \le \frac{1}{e}$ (which holds for all $x \ge 0$) for $x= \frac{k}{\alpha}$ in Eq. \eqref{eq:dis:final} leads to 
\begin{align*} 
    u\left( H_{[1, k]} \right) & \ge \left(1-\frac{1}{e^\alpha}\right) u(O).
\end{align*}
This completes the proof of theorem.

\subsubsection*{Proof of Lemma \ref{Lem:QuerySSGreedy}}
We prove this lemma in two steps. 
\newline \textbf{Step 1:} In this step, we show that $\tilde{u}$ satisfies monotonicity.
In particular, consider the compact allocation strategies $A,B \in \mathds{H}^D(\tilde{\mathcal{S}})$ and assume that $A \prec B$. We show that the revenue
extracted from each ad in $B$ is at least as much as the revenue extracted from each ad in sequence $A$. We partition the ads into two categories:
\begin{enumerate}
\item
Ads that have no budget left after running sequence $B$.
\item
Ads that still have budget after running sequence $B$.
\end{enumerate}
For the ads in the first category, sequence $B$ extracts the maximum possible budget from the ad. Therefore, for this set of ads our claim holds. For the ads that belong to the second category, we know that they still have budget available. We consider an ad
$i$ that belongs to this category and show that the revenue extracted by $B$ from this ad is at least as
much as the revenue extracted by $A$. Consider the partial configuration $(j, Y_j, \Bv^{j})$ that is being used in $B$ and not in $A$. Since advertiser $i$ never runs out of budget, not having $(j, Y_j, \Bv^{j})$ in $A$ does not change the revenue extracted from ad $i$ in serving other query types. Therefore, using sequence $B$ extracts at least as much revenue from advertiser $i$ as using sequence $A$. Since for both categories the expected revenue extracted by $B$ from each ad is higher than or equal to the
revenue extracted by $A$ from that ad, we conclude that the sequence-non-decreasing property holds.
We first prove sequence-submodularity assuming $C$ comprises of only one partial configuration, i.e., $C=(j, Y_j, \Bv^{j})$. 
As was shown in Step 1, since $A\prec B$ the remaining budget of each ad after $A$ is greater than or equal to
its remaining budget after $B$. This implies that the feasible region in finding the optimal query rewriting configuration (given in Definition \ref{def:Utility}) after running $B$ is a subset of the feasible region after running $A$,  Therefore, when we run partial configuration $(j, Y_j, \Bv^{j})$ after running $B$, we can extract less revenue compared to running it after $A$, completing the proof for $C=(j,Y_j,\Bv^{j})$. \\
For a general $C=\left((j_1, Y_{j_1}, \Bv^{j_1}), \dots, (j_k, Y_{j_k}, \Bv^{j_k}) \right)$ we then have 
\begin{align*}
&\tilde{u}(A\bot C) - \tilde{u}(A) \\
& \stackrel{(a)}{ = } \sum_{l=1}^k \tilde{u}\left( A\bot \left((j_1, Y_{j_1}, \Bv^{j_1}), \dots, (j_k, Y_{j_l}, \Bv^{j_l}) \right) \right) - \tilde{u}\left( A \bot \left((j_1, Y_{j_1}, \Bv^{j_1}), \dots, (j_k, Y_{j_l}, \Bv^{j_{l-1}}) \right) \right) \\
& \stackrel{(b)}{ \ge } \sum_{l=1}^k \tilde{u}\left((B\bot \left((j_1, Y_{j_1}, \Bv^{j_1}), \dots, (j_k, Y_{j_l}, \Bv^{j_l}) \right) \right) - \tilde{u}\left(B \bot \left((j_1, Y_{j_1}, \Bv^{j_1}), \dots, (j_k, Y_{j_l}, \Bv^{j_{l-1}}) \right) \right) \\
& \stackrel{(c)}{ = } \tilde{u}(B\bot C) - \tilde{u}(B),
\end{align*}
where (a) and (c) follow from telescopic summation (with convention for $l=0$, we let $A\bot \left((j_1, Y_{j_1}, \Bv^{j_1}), \dots, (j_k, Y_{j_l}, \Bv^{j_l}) \right)=A$) and (b) follows from the preceding proof for $C$ comprising of only one configuration and the fact that 
\begin{align*}
A \bot \left((j_1, Y_{j_1}, \Bv^{j_1}), \dots, (j_k, Y_{j_l}, \Bv^{j_{l-1}}) \right)  \prec B \bot \left((j_1, Y_{j_1}, \Bv^{j_1}), \dots, (j_k, Y_{j_l}, \Bv^{j_{l-1}}) \right), \forall l=1, \dots, k,
\end{align*}
completing the proof.
\subsubsection*{Proof of Lemma \ref{Lem:mooaw}}

In order to be able to use Algorithm \ref{alg:discrete}, at each step we need an incremental oracle to find the best partial configuration $\left(j, Y_j, \Bv^j\right)$ to be appended to the current sequence.
We claim that the marginal utility of adding a partial configuration $\left(j, Y_j, \Bv^j\right)$ is a non-decreasing submodular function
in terms of $Y_j$.
\\\textbf{Claim:} $\tilde{u}\left( \left(j, Y, \Bv^j \right)| \tilde{H} \right)$ as a function of the set $Y \subseteq \mathcal{R}$ is a non-decreasing (set) submodular function. 
\begin{proof}
 We first show the monotonicity. $\tilde{u}\left( \left(j, Y, \Bv \right)| \tilde{H} \right)$ is equal to the optimal revenue collected from serving query types $j$ by using rewrite set $Y$ when the budgets are updated after collecting the optimal revenue associated with sequence $\tilde{H}$. By expanding the set of rewrites $Y$, we have more flexibility in terms of choosing the optimal configurations and therefore the optimal collected revenue from serving queries of type $j$ increases. 
\\We next show the submodularity of $\tilde{u}\left( \left(j, Y, \Bv^j \right)| \tilde{H} \right)$. Suppose $X \subseteq Y \subseteq \mathcal{R}$ and $Z \subseteq \mathcal{R}$. We next show that
\begin{align*}
\tilde{u}\left( \left(j, X \cup Z, \Bv^j \right)| \tilde{H} \right) - \tilde{u}\left( \left(j, X , \Bv^j \right)| \tilde{H} \right) \ge \tilde{u}\left( \left(j, Y \cup Z, \Bv^j \right)| \tilde{H} \right) - \tilde{u}\left( \left(j, Y , \Bv^j \right)| \tilde{H} \right).
\end{align*}
Using the definition of the marginal utility, this inequality is equivalent to
\begin{align}\label{eq:betereki_2}
\tilde{u}\left( \left(j, X \cup Z, \Bv^j \right)\right) - \tilde{u}\left( \left(j, X , \Bv^j \right) \right) \ge \tilde{u}\left( \left(j, Y \cup Z, \Bv^j \right) \right) - \tilde{u}\left( \left(j, Y , \Bv^j \right) \right),
\end{align}
noting that the budget of advertisers are updated to the one after running sequence $\tilde{H}$. Given a partial configuration $\left(j, Y , \Bv^j \right)$, we can use the ads in the set $\cup_{r \in Y} W_r$ to serve queries of type $j$. Using Definition \ref{def:Utility}, we can compute $\tilde{u}\left( \left(j, Y , \Bv^j \right)\right)$ by using a sequence of ads obtained as follows. We sort the payments of the ads in $\cup_{r \in Y} W_r$ and then include the ad with the top payment until we exhaust its entire budget, we then include the ad with the second top payment and continue this process. We denote this sequence of ads by $\mathrm{seq}(\cup_{r \in Y} W_r, \Bv^{j})$. The revenue function $\tilde{u}\left( \left(j, Y , \Bv^j \right)\right)$ is given by 
\begin{align*}
\tilde{u}\left( \left(j, Y , \Bv^j \right)\right)= \sum_{l=1}^{T q_j}  p_{\mathrm{seq}_l(\cup_{r \in Y} W_r, \Bv^{j}), j} 
\end{align*}
where $\mathrm{seq}_l(\cup_{r \in Y} W_r, \Bv^{j})$ denotes the $l$-th element of the sequence $\mathrm{seq}(\cup_{r \in Y} W_r, \Bv^{j})$. Using this notation we obtain 
\begin{align}\label{eq:betereki}
& \tilde{u}\left( \left(j, Y \cup Z, \Bv^j \right)\right) - \tilde{u}\left( \left(j, Y , \Bv^j \right) \right) \nonumber \\
& \stackrel{(a)}{=} \sum_{l=1}^{T q_j} p_{\mathrm{seq}_l(\cup_{r \in Y \cup Z} W_r, \Bv^{j}), j} - \sum_{l=1}^{T q_j} p_{\mathrm{seq}_l(\cup_{r \in Y} W_r, \Bv^{j}), j} \\
& \stackrel{(b)}{=} \sum_{l=1}^{T q_j} p_{\mathrm{seq}_l(\cup_{r \in Y \cup Z} W_r, \Bv^{j}), j} \mathbf{1}\{\mathrm{seq}_l(\cup_{r \in Y \cup Z} W_r, \Bv^{j}) \in \cup_{r \in Y} W_r\} \nonumber \\
& + \sum_{l=1}^{T q_j} p_{\mathrm{seq}_l(\cup_{r \in Y \cup Z} W_r, \Bv^{j}), j} \mathbf{1}\{\mathrm{seq}_l(\cup_{r \in Y \cup Z} W_r, \Bv^{j}) \not\in \cup_{r \in Y} W_r\} - \sum_{l=1}^{T q_j} p_{\mathrm{seq}_l(\cup_{r \in Y} W_r, \Bv^{j}), j}  \nonumber \\
& \stackrel{(c)}{=} \sum_{l=1}^{T q_j- K} p_{\mathrm{seq}_l(\cup_{r \in Y} W_r, \Bv^{j}), j} + \sum_{l=1}^{K} p_{\mathrm{seq}_l\left(\left( \cup_{r \in Y \cup Z} W_r \right) \setminus \left(\cup_{r \in Y} W_r \right), \Bv^{j} \right), j} - \sum_{l=1}^{T q_j} p_{\mathrm{seq}_l(\cup_{r \in Y} W_r, \Bv^{j}), j} \nonumber \\
& = \sum_{l=1}^{K} p_{\mathrm{seq}_l\left(\left( \cup_{r \in Y \cup Z} W_r \right) \setminus \left(\cup_{r \in Y} W_r \right), \Bv^{j} \right), j} - \sum_{l=Tq_j-K}^{T q_j} p_{\mathrm{seq}_l(\cup_{r \in Y} W_r, \Bv^{j}), j}.
\end{align}
where $K= \sum_{l=1}^{T q_j} \mathbf{1}\{\mathrm{seq}_l(\cup_{r \in Y \cup Z} W_r, \Bv^{j}) \not\in \cup_{r \in Y} W_r\}$. Note that (a) follows from the definition of $\mathrm{seq}(\cdot, \cdot)$, (b) simply follows because $\mathrm{seq}_l(\cup_{r \in Y \cup Z} W_r, \Bv^{j})$ either belongs to $\cup_{r \in Y} W_r$ or not, and (c) follows because the top $T q_j -K$ ads have appeared in the top $T q_j$ ads of $\mathrm{seq}_l(\cup_{r \in Y } W_r, \Bv^{j})$. 

We next show inequality \eqref{eq:betereki_2}. We can write 
\begin{align*}
& \tilde{u}\left( \left(j, X \cup Z, \Bv^j \right)\right) - \tilde{u}\left( \left(j, X , \Bv^j \right) \right) \\
& \stackrel{(a)}{\ge} \sum_{l=1}^{K} p_{\mathrm{seq}_l\left(\left( \cup_{r \in Y \cup Z} W_r \right) \setminus \left(\cup_{r \in Y} W_r \right), \Bv^{j} \right), j} - \sum_{l=Tq_j-K}^{T q_j} p_{\mathrm{seq}_l(\cup_{r \in X} W_r, \Bv^{j}), j} \\
& \stackrel{(b)}{\ge} \sum_{l=1}^{K} p_{\mathrm{seq}_l\left(\left( \cup_{r \in Y \cup Z} W_r \right) \setminus \left(\cup_{r \in Y} W_r \right), \Bv^{j} \right), j} - \sum_{l=Tq_j-K}^{T q_j} p_{\mathrm{seq}_l(\cup_{r \in Y} W_r, \Bv^{j}), j} \\
&\stackrel{(c)}{=} \tilde{u}\left( \left(j, Y \cup Z, \Bv^j \right)\right) - \tilde{u}\left( \left(j, Y , \Bv^j \right) \right),
\end{align*}
where (a) follows from the fact that any ad $l$ that belongs $\left( \cup_{r \in Y \cup Z} W_r \right) \setminus \left(\cup_{r \in Y} W_r \right)$ belongs to the set $\left( \cup_{r \in X \cup Z} W_r \right) \setminus \left(\cup_{r \in X} W_r \right)$ as well, (b) follows from the fact that there are more ads in the set $\cup_{r \in Y} W_r$ compared to the set $\cup_{r \in X} W_r$ and therefore the bottom $K$ ads (among the top $T q_j$ ads) have larger payments when we use $Y$, and (c) directly follows from Eq. \eqref{eq:betereki}. 
%
%
This completes the proof of the claim. 
\end{proof}
Using the claim, for each $j$ the greedy algorithm finds $Y_j$ that obtains $1- {1}/{e}$ of the optimal set of rewrites subject to cardinality constraint $|Y_j| \le k$. 
The greedy algorithm starts from an empty $Y_j$ and adds the rewrite that increases the
marginal utility the most, until $k$ rewrites have been added. 

The algorithm then selects among all possible query types
$j$, the one for which $\left(j, Y_j, \Bv^j\right)$ has the highest marginal utility and appends it to the current
sequence of configurations. Now we can use Theorem \ref{thm:dis} and Corollary \ref{thm:dis:fin} with $\alpha= 1- {1}/{e}$ which guarantees that the approximation ratio of the overall algorithm is
$1- {1}/{e^{1-\frac{1}{e}}}$. Therefore, letting $H^*$ be the output of Algorithm \ref{alg:rew}, we obtain 
\begin{align*}
\tilde{u}(H^*) \ge \left(1- \frac{1}{e^{1- \frac{1}{e}}} \right) \max_{\tilde{H} \in \mathds{H}^D(\mathcal{S}^p)} \tilde{u}(\tilde{H}).
\end{align*}
This completes the proof. 

\subsubsection*{Proof of Theorem \ref{pro:Equivalent}}
Before proving this theorem, we introduce a few notations. We let $\tilde{O}$ be the optimal compact allocation strategy in $\mathds{H}^D(\tilde{S})$ together with sets $Y_j$ for all $j \in \mathcal{Q}$. We also let $H^*$ be the output of Algorithm \ref{alg:rew} and ${H^*}^q$ be its corresponding query rewriting allocation strategy. Finally, we let $O^q$ denote the optimal query rewriting allocation strategy of the original query rewriting problem. 

We now proceed with proving this theorem in three steps.  
\\ \textbf{Step 1:} In the first step, we establish the relation between $\tilde{u}(\tilde{O})$ and $u({H^*}^q)$. In particular, we show
\begin{align*}
u({H^*}^q) \ge \left(1- \frac{1}{e^{1- \frac{1}{e}}} \right) \tilde{u}(\tilde{O})- \left( \frac{\max_{i \in \mathcal{A}, j \in \mathcal{Q}} p_{ij}}{ \min_{i \in \mathcal{A}} B_i} \right) \tilde{u}(\tilde{O}).
\end{align*}
\emph{Proof of Step 1:} If we could have fractional allocation of ads to queries, then the allocation strategy ${H^*}^q$ would have the same revenue as that of $H^*$. Here, we use a similar argument to that of Lemma \ref{Lem:jaDd_1} to show the relation between $\tilde{u}(H^*)$ and $u({H^*}^q)$. In particular, we have 
\begin{align}\label{eq:tomp_2_2}
u({H^*}^q) \ge \tilde{u} (H^*) - \sum_{i \in \mathcal{A},~ i\text{'s budget in } H^* \text{ is exhausted }} p_{ij},
\end{align}
We also have 
\begin{align}\label{eq:tomp_4_2}
\sum_{i \in \mathcal{A},~ i\text{'s budget in } H^* \text{ is exhausted }} p_{ij} & \le \left( \frac{\max_{i \in \mathcal{A}, j \in \mathcal{Q}} p_{ij}}{ \min_{i \in \mathcal{A}} B_i} \right) \sum_{i \in \mathcal{A},~ i\text{'s budget in } H^* \text{ is exhausted }} B_i  \nonumber \\
& \le  \left(\frac{\max_{i \in \mathcal{A}, j \in \mathcal{Q}} p_{ij}}{ \min_{i \in \mathcal{A}} B_i} \right) \tilde{u}(H^*) \le \left( \frac{\max_{i \in \mathcal{A}, j \in \mathcal{Q}} p_{ij}}{ \min_{i \in \mathcal{A}} B_i} \right) \tilde{u}(\tilde{O}). 
\end{align}
Invoking \eqref{eq:tomp_4_2} in \eqref{eq:tomp_2_2}, leads to 
\begin{align}\label{eq:tomp_5_2}
u({H^*}^q) & \ge \tilde{u} (H^*)- \left( \frac{\max_{i \in \mathcal{A}, j \in \mathcal{Q}} p_{ij}}{ \min_{i \in \mathcal{A}} B_i} \right) \tilde{u}(\tilde{O}) .
\end{align}
On the other hand, using Lemma \ref{Lem:mooaw} we obtain 
\begin{align}\label{eq:tomp_6_2}
\tilde{u}(H^*) \ge \left(1- \frac{1}{e^{1- \frac{1}{e}}} \right) \tilde{u}(\tilde{O}).
\end{align}
Using \eqref{eq:tomp_6_2} in \eqref{eq:tomp_5_2}, leads to 
\begin{align}\label{eq:tomp_7_2}
u({H^*}^q) \ge \left(1- \frac{1}{e^{1- \frac{1}{e}}} \right) \tilde{u}(\tilde{O})- \left( \frac{\max_{i \in \mathcal{A}, j \in \mathcal{Q}} p_{ij}}{ \min_{i \in \mathcal{A}} B_i} \right) \tilde{u}(\tilde{O}).
\end{align}
This completes the proof of the first step. 
\\ \textbf{Step 2:} In this step, we show the connection between $\tilde{u}(\tilde{O})$ and $u(O^q)$. In particular, we show that $\tilde{u}(\tilde{O}) \ge u(O^q)$. 

We let $\hat{O}$ be the optimal offline allocation over all possible ordering of incoming queries. That is is finding $\hat{O}$ we assume all the queries have arrived with a particular order and then find the optimal query rewriting allocation strategy for this ordering of queries. Clearly, we have $u(\hat{O}) \ge u(O^q)$. We next show that $\hat{O}$ defines a corresponding compact allocation strategy whose revenue function is equal to $u(\hat{O})$. In particular, we let $B_{ij}$ be the budget that ad $i$ has consumed in serving query $j$ using $\hat{O}$. We also let $Y_j$ be the set of rewrites used in $\hat{O}$. For the compact allocation strategy
\begin{align*}
\bar{O}=((1, Y_1, \Bv^{1}), \cdots, (n, Y_n, \Bv^{n})),
\end{align*} 
we have $\tilde{u}(\bar{O})= u(\hat{O})$. Therefore, we obtain 
\begin{align*}
\tilde{u}(\tilde{O}) \ge \tilde{u}(\bar{O})= u(\hat{O}) \ge u(O^q).
\end{align*}
 Combining the first and the second steps, we obtain 
 \begin{align*}
 u({H^*}^q) \ge \left(1- \frac{1}{e^{1- \frac{1}{e}}} \right) u(O^q)- \left( \frac{\max_{i \in \mathcal{A}, j \in \mathcal{Q}} p_{ij}}{ \min_{i \in \mathcal{A}} B_i} \right) u(O^q),
 \end{align*}
 which completes the proof.

\subsubsection*{Generalization of Query Rewriting Analysis to $d >1 $}\label{App:General}
In order to generalize the analysis to a setting with $d >1$, we first introduce the generalization of revenue function given in Definition \ref{def:Utility}. We then show that given any compact allocation strategy $\tilde{H}$, there exists a corresponding query rewriting allocation strategy whose expected revenue is equal $\tilde{u}(\tilde{H})$ (similar to the argument used in the proof of Theorem \ref{pro:Equivalent}, this is assuming for the ads that run out of budget, the ad allocator can show a fraction of the ad and extract its remaining budget). 
\\\textbf{Definition of revenue function $\tilde{u}$:}
For any compact allocation  strategy $\tilde{H}$, its revenue function denoted by $\tilde{u}(\tilde{H})$ is the revenue collected by sequentially running the partial configurations specified by sequence $\tilde{H}$ and extract the optimal revenue given the budget constraints. Formally, we let $\tilde{u}(\emptyset)$ be zero and $\tilde{B}_i$ be the current budget of advertiser $i$ initialized to $B_i$. Suppose $( j, Y_{j}, \Bv^j)$ is the current element of $\tilde{H}$ (with the order specified by $\tilde{H}$). The current budget limit of advertiser $i$ is the minimum of $\tilde{B}_i$ and $B_{ij}$.

We first solve the following linear programming and add its optimal value to the current value of $\tilde{u}(\tilde{H})$. 
\begin{align}
\max_{x_1, \dots, x_n} & \sum_{i=1}^m x_i p_{ij} \label{Eq:GeneralQ_0} \\
& x_i p_{ij} \le \min\{\tilde{B}_i, B_{ij}\} , \quad \forall i \in \bigcup_{r \in Y_j} W_r, \label{Eq:GeneralQ}\\
& x_i  \le T q_j , \quad \forall i \in \bigcup_{r \in Y_j} W_r, \label{Eq:GeneralQ_3}\\
& \sum_{i=1}^m x_i \le d T q_j. \label{Eq:GeneralQ_2}
\end{align}
In this linear programming, $x_i$ denotes the number of ads from advertiser $i$ shown to queries of type $j$ (note that $x_i$ may be non-integer in which case, the ad allocator shows the ad for a fraction of time). The constraints given in inequality \eqref{Eq:GeneralQ} is to capture the fact that the allowed budget of advertiser $i$ for query types $j$ is $B_i$. The constraint given in inequality \eqref{Eq:GeneralQ_3} captures that we do not have more than $T q_j$ queries of type $j$ and for any query we cannot an ad in more than one slot (among the available $d$ slots). Finally, the constraint given in Eq. \eqref{Eq:GeneralQ_2} captures the fact that there are $T q_j$ queries of type $j$ and for each one of them the ad allocator can show at most $d$ ads. 
 
We then update the budgets $B_i$ by subtracting the optimal $x_i$. Finally, we proceed to the next partial configuration in the sequence $\tilde{H}$. The revenue function $\tilde{u}(\tilde{H})$ is the revenue obtained at the end of this procedure.

\begin{figure}[t]
\centering
    \includegraphics[width=0.75\textwidth]{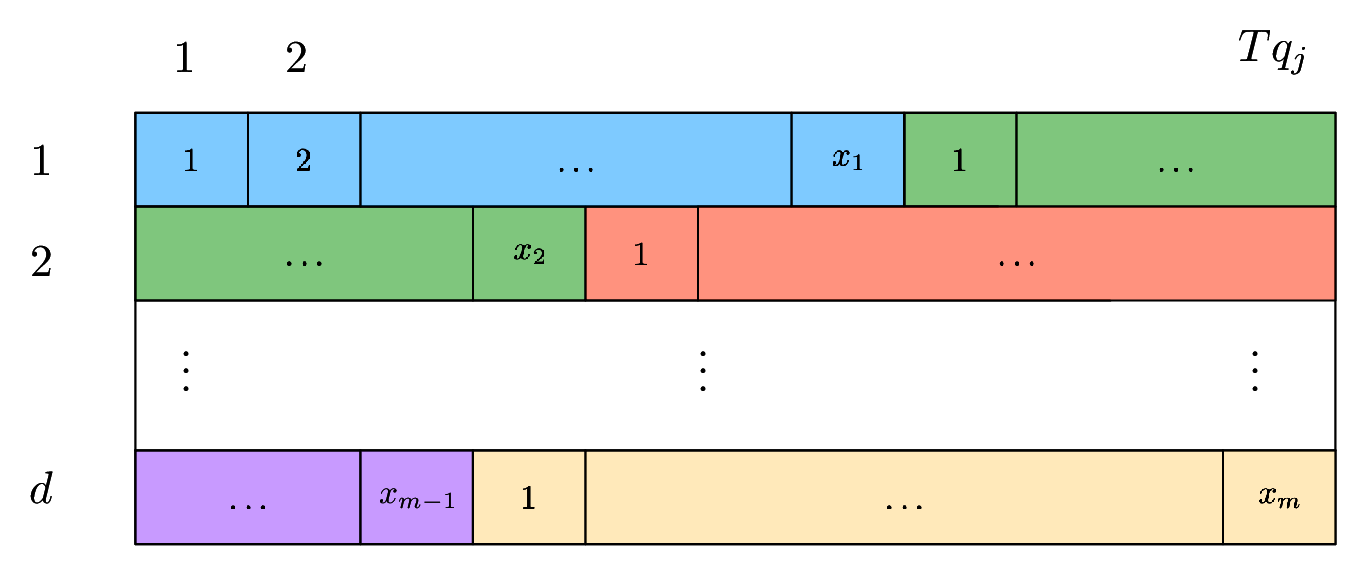}
\caption{Implementation of a solution of problem \eqref{Eq:GeneralQ_0}. We start from the first row and serve ad $1$ for $x_1$ queries and then serve the next add until we serve all the ads.}
    \label{fig:Marmoolak}
\end{figure}

We next show that the solution of the linear programming given in Eq. \eqref{Eq:GeneralQ_0} is implementable. That is we show how the solution of problem \eqref{Eq:GeneralQ_0} specifies a query rewriting allocation strategy whose collected revenue is the same as the objective of the optimal solution of problem \eqref{Eq:GeneralQ_0}. Note that the main challenge is that for any query type $j$, we can only show $d$ ads which need to be distinct from each other and it is not clear whether a solution of problem \eqref{Eq:GeneralQ_0} is implementable in view of these constraints. We first show an example illustrating the aforementioned challenge.
\begin{example}
\textup{
Suppose we have one query type, denoted by $1$, which arrive $10$ times and we have $3$ ads, denoted by $\{1, 2, 3\}$, each of them with budget $10$. Also suppose we have two ad slots per query (i.e., $d=2$) and the payments are $p_{11}=1$ and $p_{21}=p_{31}=2$. For this example, the solution of problem \eqref{Eq:GeneralQ_0} becomes $x_1=10$, $x_{2}=x_{3}=5$ with optimal revenue $30$. Now if we greedily allocate ads to the slots, then we need to serve the first $5$ queries with ads $\{2,3\}$ which have the maximum payment, then for the remaining $5$ queries we can only show one ad of type $1$ (this is because we cannot show ad $1$ in more than one of slots available for each query). Therefore, the overall revenue of a greedy implementation becomes $25$ which is not the same as the optimal solution of problem \eqref{Eq:GeneralQ_0}. However, we serve the first $5$ queries with ads $\{1, 2\}$ and last $5$ queries with ads $\{1,3\}$, then the resulting revenue becomes $30$. This example illustrates that a greedy allocation of ads to queries does not necessarily lead to the optimal revenue. 
}
\end{example}
We next show how we can implement the optimal solution of problem \eqref{Eq:GeneralQ_0}. Suppose $x_1, \dots, x_m$ denote the optimal solution. Without loss of generality assume $x_1 \ge \dots, \ge x_m$. The implementation is as follows. We allocate ad $1$ to the first ad slot of the first $x_1$ queries. We then allocate ad $2$ to the remaining first ad slots until we reach the last query (i.e., the $T q_j$-th query). We then allocate the remaining ads of type $2$ to the second ad slots of the initial queries. We continue this procedure until all ads are allocated. First note that all ads will be allocated because of the constraint $\sum_{i=1}^m x_i \le dTq_j$. Moreover, the ads shown for each query are distinct. This is because of the way we fill out the slots and the fact that $x_i \le T q_j$ for all $i=1, \dots, m$. Figure \ref{fig:Marmoolak} illustrates this procedure. 

The rest of the analysis is identical to that of Lemma \ref{Lem:mooaw} and Theorem \ref{pro:Equivalent}.  


\bibliographystyle{plainnat}
\bibliography{References}

\end{document}